\documentclass[pre,twocolumn,nofootinbib,twoside,showpacs,superscriptaddress,tightenlines,floatfix,aps]{revtex4-2}
\usepackage{dcolumn}
\usepackage{lipsum}
\usepackage{graphicx,amssymb,amsmath,epsf,bm}
\usepackage[utf8x]{inputenc}
\usepackage[T1]{fontenc}
\usepackage[normalem]{ulem}
\usepackage{color}
\usepackage{bm}
\usepackage{tikz}
\usepackage{mathtools}
\usepackage{dsfont}
\usepackage{nicefrac}
\usepackage{hyperref}
\hypersetup{
    allcolors=blue,
    colorlinks=true,
    bookmarks=true,
    pdfpagemode=FullScreen,
}
\usepackage{ulem}
\usepackage{tikz}
\usepackage[compat=1.1.0]{tikz-feynman}
\usepackage{graphicx} 
\usepackage{bm}
\usepackage{physics}
\usepackage{txfonts}
\usepackage{mathrsfs}
\usepackage{amsmath}
\usepackage{feynmf}
\usepackage{comment}
\usepackage{xcolor}
\usepackage{upgreek}

\usepackage{hyperref}
\hypersetup{
    allcolors=blue,
    colorlinks=true,
    bookmarks=true,
    pdfpagemode=FullScreen,
}

\definecolor{nred}{RGB}{224,0,0}
\definecolor{nblue}  {RGB}{28,130,185}
\definecolor{dgreen} {RGB}{38,238,21}
\definecolor{norange}{RGB}{230,120,20}

\renewcommand{\Tr}{{\rm Tr}}

\begin{document}

\title{Critical Probability Distributions of the order parameter at two loops II: $O(n)$ universality class }

\author{Sankarshan Sahu}
\affiliation{Sorbonne Universit\'e, CNRS, Laboratoire de Physique
Th\'eorique de la Mati\`ere Condens\'ee, LPTMC, 75005 Paris, France}
\date{\today}
\begin{abstract}
We show how to compute the probability distributions of the order parameter of the $O(n)$ model at  two-loop order of perturbation theory  generalizing the methods developed for computing the same in case of the Ising model \cite{Sahu:2025bkp}. We show that even for the $O(n)$ model, there exists not one but a family of these probability distribution functions indexed by $\zeta$ which is the ratio of system size $L$ to the bulk correlation length $\xi_{\infty}$. We also compare these PDFs  to the Monte-Carlo simulations and the existing FRG results \cite{Rancon:2025bjf} for the $O(2)$ and $O(3)$ models. 
\end{abstract}


\maketitle

\section{Introduction}
 In \cite{eisenriegler1987helmholtz} and in our previous work \cite{sahu2024generalization}, the computation of the Probability Distribution Functions (PDFs) of the order-parameter (or more precisely the normalized total spin)  at criticality has been achieved perturbatively at one-loop for both $O(n)$ as well as Ising model in the $\epsilon=4-d$ expansion. In the companion paper (hereafter referred to as I) \cite{Sahu:2025bkp}, we have computed these critical PDFs for the Ising model at two loops. The goal of the current article is to generalize these results to the $O(n)$ models at two loops.  Just as at one loop and at two loops (for the Ising case), we end up with not one but a family of these PDFs characterized by $\zeta$ with $\zeta$ being the ratio of system size $L$ to bulk correlation length $\xi_{\infty}$. Although there had been previous attempts to compute such PDFs for the $O(n)$ model at criticality \cite{PhysRevLett.77.3641, Hilfer1995, doi:10.1142/S0217979298000703, Esser:1995id, PhysRevE.58.2902} for $\zeta=0$, our aim is to build a systematic perturbative approach towards computing the entire family of PDFs indexed by $\zeta$.

\section{Field Theoretic Formalism}
We are interested in computing the PDFs of the normalized total spin in the $O(n)$ model.
\begin{equation}\label{P}
    \hat{P}(\hat{s}_{i}=s_{i}) \propto\int ~\prod_{i}D\hat{\phi}_{i}~~\delta(\hat{s}_{i}-s_{i})\exp\left(-\int\mathcal{H}[\hat{\phi}_{i}]\right),
\end{equation}
where:
\begin{equation}
    \hat{s}_{i} =\frac{1}{L^{d}}\int_{x}~\hat{\phi}_{i}(x)
\end{equation}
with $i=1,\cdots,n$, $\int_{x}=\int d^{d}x$, and the Hamiltonian being given by:
\begin{equation}\label{Hamil}
    \mathcal{H}[\hat{\phi}_{i}]=\int_{x}\frac{Z_{1}}{2}\left(\nabla\hat{\phi}_{i}(x)\right)^2+\frac{1}{2}Z_{2}\,t\,({\hat{\phi}^{2}_{i}}(x))+\frac{1}{4!}Z_{4}\,g\,({\hat{\phi}}_{i}^2(x))^2.
\end{equation}
(Einstein summation convention is assumed throughout).\\
 Exploiting rotational invariance one can further argue that $ P(\hat{s}_{i}=s_{i})$ 
is only a function of $\sum_{i}s^{2}_{i}=s^2$. Just as in I,  we choose to parameterize the theory in terms of the renormalized temperature difference $t$, the renormalized coupling constant $g$ and the renormalized fields $\hat{\phi}_{i}$ defined at an infrared scale $\mu$ instead of the bare quantities $t_{0}$ $g_{0}$ and $\hat{\phi}^{\Lambda}_{i}$ defined at an ultraviolet scale $\Lambda$ with $\Lambda\gg\mu$. These quantities are hence related via the counter-terms:
\begin{equation}
    t_{0}=Z'_{2}t\ , \ \ g_{0}=Z'_{4}g\ , \ \ \hat{\phi}^{\Lambda}_{i}=Z^{1/2}_{1}\hat{\phi}_{i},
\end{equation}
with 
\begin{equation}
    Z_{2}=Z_{1}Z'_{2}\ , \ \ Z_{4}=Z^{2}_{1}Z'_{4}\ .
\end{equation}
 In the following, we use dimensional regularization, where $Z_{1}$, $Z_{2}$ and $Z_{4}$ are counter-terms introduced in accordance with the MS scheme just as in I.

Replacing the delta function in Eq.~\eqref{P} by a sharply peaked Gaussian i.e. $\delta(z)\propto\exp\left(-\frac{M^2 z^2}{2}\right)$ with $M\rightarrow\infty$ as in \cite{balog2022critical, sahu2024generalization}, the PDF can be interpreted as the partition function (with $\mathcal{N}$ a normalization constant):
\begin{equation}
 \mathcal{Z}_{M,s}[\mathbf{h}] =\mathcal{N}\int D\hat{\phi}\exp\left(-\mathcal{H}_{M,s}(\hat{\phi}_{i}(x))+\int_x h_{i}(x)\hat{\phi}_{i}(x)\right),
\end{equation}
  at vanishing magnetic field $\mathbf{h}=0$ (here $\mathbf{h}$ is a vector with $\mathbf{h} =\{h_{i}\}$)  of a system with  Hamiltonian
 \begin{equation}
     \mathcal{H}_{M,s}\left(\hat{\phi}_{i}(x)\right)=\mathcal{H}\left(\hat{\phi}_{i}(x)\right)+ \frac{M^2}{2}\left(\int_{x}(\hat{\phi}_{i}(x)-s_{i})\right)^2.
 \end{equation} 
 The modified Gibbs free energy $\Gamma_{M}[\bm{\phi}]$ is defined by \cite{Rancon:2025bjf}:
 \begin{equation}\label{PE1}
    \Gamma_{M}[\bm{\phi}]+\log{\mathcal{Z}_{M,s}[\mathbf{h}]} =  h_{i}.\phi_{i}-\frac{M^2}{2}\left(\int_{x}({\phi}_{i}(x)-s_{i})\right)^2,
\end{equation}
with
\begin{equation}\label{phi}
     \phi_{i}(x) = \frac{\delta \log{\mathcal{Z}_{M, s}[\mathbf{h}]}}{\delta h_{i}(x)}, 
 \end{equation}
and $h_{i}.\phi_{i}=\int_{x} h_{i}(x)\phi_{i}(x)$. \\
 
 One can also subsequently define:
 \begin{equation}
     W_{M,s}[\mathbf{h}]=\log{\mathcal{Z}_{M,s}[\mathbf{h}]}.
 \end{equation}
 Using tricks similar to \cite{balog2022critical, sahu2024generalization, Sahu:2025bkp} , one can hence show :
 \begin{align}
     \lim_{M\rightarrow\infty} e^{-\Gamma_{M}[\phi_{i}=s_{i}]} & = \mathcal{N}\int ~\prod_{i}D\hat{\phi}_{i}~~\delta(\hat{s}_{i}-s_{i})\exp\left(-\int\mathcal{H}[\hat{\phi}_{i}]\right)\nonumber\\
     & \propto \hat{P}(\hat{s}_{i}=s_{i}).
 \end{align}
 We define the rate function $I(s)$ in the following way:
 \begin{equation}
     \hat{P}(\hat{s}_{i}=s_{i})\propto e^{-L^{d}\hat{I}(s,\, \xi_{\infty},\,  L)},
 \end{equation}
 and thus :
\begin{equation}\label{i(s)}
    \lim_{M\rightarrow\infty}\Gamma_{M}[\phi_{i}(x)=s_{i}] =L^{d}\hat{I}(s,\, \xi_{\infty},\,  L),
\end{equation}
 with $s^{2}=\sum_{i}s^{2}_{i}$.
\section{The  $O(n)$ Rate Function at two loops}\label{3}
As shown in \cite{sahu2024generalization}, the rate function at one-loop can be easily obtained by expanding $\log{\mathcal{Z}_{M, s}[h]}$ around the mean-field configuration $\hat{\phi}^{(0)}_{i}$, defined by:
\begin{equation}
    \frac{\delta \mathcal{H}_{M, s_{i}}}{\delta \hat{\phi}_{i}}|_{\hat{\phi}_{i}=\hat{\phi}^{(0)}_{i}}=h_{i}.
\end{equation}
Doing so, one thus finds:
\begin{equation}
    \Gamma_{M}[\bm{\phi}] = \mathcal{H}[\bm{\phi}]+\frac{1}{2}\Tr\log\mathcal{H}^{(2)}_{M, s}-\log{[\mathcal{N}]},
\end{equation}
 (here $\Gamma_{M}[\bm {\phi}]$ has been evaluated at a constant field).
where $\log[\mathcal{N}]$ has  been chosen such that it is equal to the value of $\mathcal{H}[\bm{\phi}]+\Tr\log\mathcal{H}^{(2)}_{M, s}[\bm{\phi}]$ computed at vanishing field and vanishing mass (i.e. at $T=T_{c}$ with $T_{c}$ the critical temperature) with an addition of constant-infinite pole counter-terms to cancel out the divergence divergences coming out from the one-loop term (in this case and higher order loop terms corresponding to the loop order up-to which we perform the perturbative expansion). Using arguments from \cite{jackiw1974functional} and I\cite{Sahu:2025bkp}, one can show that $\Gamma_M[\bm{\phi}]$ re-sums the 1-Particle Irreducible diagrams (or the 1PIs) appearing in the perturbative expansion of $W^{(2)}_{M,s}[\mathbf{h}]$, where\footnote{$W^{(2)}_{M,s}[\mathbf{h}]$ is not the double derivative of $W_{M,s}[\mathbf{h}]$ but rather its loop expansion as seen in Eq.~\eqref{Z2 M}.},:

\begin{align}
W^{(2)}_{M,s}[\mathbf{h}]=\log{Z^{(2)}_{M,s}[\mathbf{h}]},
\end{align}
with 
\begin{align}\label{Z2 M}
& Z^{(2)}_{M,s}[\mathbf{h}]\nonumber\\
& =\frac{\int \mathcal{D\hat{\phi}}\exp{-\left(\mathcal{H}_{M,s}[\hat{\phi}+\hat{\phi}^{(0)}_{i}]- \int_{x} h_{i}(x)\hat{\phi}^{(0)}_{i}(x)-\mathcal{H}_{M,s}[\hat{\phi}]\right)}}{\int \mathcal{D\hat{\phi}}\exp{-\left(\int\hat{\phi_{i}}(x)\frac{\delta^2\mathcal{H}_{M,s}(x,y)}{\delta\hat{\phi}_{i}\delta\hat{\phi}_{j}}|_{\hat{\phi}_{i}=\hat{\phi}^{(0)}_{i}}\hat{\phi}_{j}(y)\right)}}.
\end{align}
 One can easily compute these 1PIs by computing the propagator of the theory and thereby using Wick's Theorem. The propagator of the theory is given by :
\begin{equation}\label{prop}
\mathcal{D}^{-1}_{ij}(x, y) =   \frac{\delta^2\mathcal{H}_{M,s}}{\delta\hat{\phi}_{i}(x)\delta\hat{\phi}_{j}(y)}
\end{equation}
In Fourier space, Eq.~\eqref{prop} becomes:
\begin{equation}\label{prop1}
    \mathcal{D}^{-1}_{ij}(q)=\left(Z_{1}q^2+Z_{2}t+Z_{4}\frac{g\phi^{2}}{6}+M^2\delta_{q, 0}\right)\delta_{ij}+\frac{Z_{4}g}{3}\phi_{i}\phi_{j} .
\end{equation}
Notice that the propagator in Eq.~\eqref{prop1} is not the infinite volume propagator of the $O(n)$  model. It differs from the 'usual' propagator by the $M^2\delta_{q, 0}\delta_{ij}$ term.  Thus at two loops, the modified effective potential is given by (i.e. $\Gamma_{M}[\bm {\phi}]$ evaluated at a constant field):
    \begin{align}\label{Gammarate M}
 \Gamma_{M}[\bm{\phi}] & =  \mathcal{H}[\bm{\phi}]+\frac{1}{2}\Tr\log{\mathcal{H}^{(2)}_{M,s}}[\bm{\phi}]+\frac{1}{24}(\mathcal{I}_{3}+\mathcal{I}_{4}+\mathcal{I}_{5})\nonumber\\
 &-\frac{1}{36}(\mathcal{I}_{1}+\mathcal{I}_{2})-\log[\mathcal{N}].
 \end{align}
 with :\\

\begin{tikzpicture}
    \begin{feynman}
     \vertex(a);
     \vertex [right =of a] (b); 
     \vertex [left=0.3em of a] {$\mathcal{I}_{1}=
3$}; 
\vertex [right=0.5em of b];
 \diagram{      
            (a) -- (b); 
            (a) --(b);
            (a) --[half left](b);
            (a) --[half right](b);
            
        };

    \end{feynman}
 \end{tikzpicture}\\
 \begin{tikzpicture}
\begin{feynman}    
    
    \vertex(c);
     \vertex [right =of c] (d); 
     \vertex [left=0.3em of c] {$\mathcal{I}_{2}=
(n-1)$}; 
\vertex [right=0.5em of d];
 \diagram{      
            (c) -- (d); 
            (c) --(d);
            (c) --[half left, scalar](d);
            (a) --[half right, scalar](d);
            
        };

    \end{feynman}
\end{tikzpicture}\\
\begin{tikzpicture}

    \begin{feynman}
     \vertex(a);
     \vertex [right =of a] (b); 
      \vertex [right =of b] (c);
      \vertex [left=0.2em of a] {$\mathcal{I}_{3}=3$};
      
        \diagram{
            
            (a)--[half right](b);
            (a) --[half left](b);
            (b) --[half left](c);
            (b) --[half right](c);
        };

    \end{feynman}
    
   \end{tikzpicture}\\
   \begin{tikzpicture}

    \begin{feynman}
     \vertex(a);
     \vertex [right =of a] (b); 
      \vertex [right =of b] (c);
      \vertex [left=0.2em of a] {$\mathcal{I}_{4}=(n^2-1)$};
      
        \diagram{
            
            (a)--[half right, scalar](b);
            (a) --[half left, scalar](b);
            (b) --[half left, scalar](c);
            (b) --[half right, scalar](c);
        };

    \end{feynman}
    
   \end{tikzpicture}\\
\begin{tikzpicture}
    \begin{feynman}
     \vertex(a);
     \vertex [right =of a] (b); 
      \vertex [right =of b] (c);
      \vertex [left=0.2em of a] {$\mathcal{I}_{5}=2(n-1)$};
      
        \diagram{
            
            (a)--[half right](b);
            (a) --[half left](b);
            (b) --[half left, scalar](c);
            (b) --[half right, scalar](c);
        };

    \end{feynman}
\end{tikzpicture}\\  
where: \\

 \begin{tikzpicture}[scale=0.2]
    \begin{feynman}
     \vertex(a);
     \vertex [right =of a](b); 
     \vertex [left=0.3em of a]{$\mathcal{D}_{1}^{-1}=$}; 
\vertex [right=0.4em of b] {$=Z_{1}q^2+M^2\delta_{q, 0}+Z_{2}t+Z_{4}{g\phi ^2}/{6},
$};
 \diagram{      
            (a) -- [scalar](b); 
            
        };

    \end{feynman}
\end{tikzpicture}

and\\

 \begin{tikzpicture}[scale=0.2]
    \begin{feynman}
     \vertex(a);
     \vertex [right =of a] (b); 
     \vertex [left=0.3em of a]{$\mathcal{D}_{2}^{-1}=$}; 
\vertex [right=0.4em of b] {$= Z_{1}q^2+M^2\delta_{q, 0}+Z_{2}t+Z_{4}{g\phi ^2}/{2}.
$};
 \diagram{      
            (a) -- (b); 
            
        };

    \end{feynman}
\end{tikzpicture}
\\

As is the usual case with $O(n)$ models, the field theory can be treated in terms of two propagators the longitudinal one $\mathcal{D}^{-1}_{1}$ and the transverse one $\mathcal{D}^{-1}_{2}$. Also, we note that there are two kinds of vertices in the theory: the 3- and the 4-point vertex: all the 3-point vertices are given by $g\bm{\phi}/6$ while all the 4-point vertices are given by $g/24$ .\\
Using Eq.~\eqref{Gammarate M}, in the limit $M\rightarrow\infty$ one thus obtains: 
\begin{widetext}
 \begin{align}\label{e1}
      \lim_{M\rightarrow\infty}\Gamma_M[\phi_{i}(x)=s_{i}]& = L^d\biggl(\frac{1}{2}Z_{2}t s^2+\frac{1}{4!}Z_{4}g s^4+\frac{1}{2L^d}\sum_{\vec{q}\neq 0}\log\left(1+\frac{m_{2}^2}{\vec{q}^2}\right)+\frac{(n-1)}{2L^d}\sum_{\vec{q}\neq 0}\log\left(1+\frac{m_{1}^2}{\vec{q}^2}\right)\nonumber\\
 &+\frac{g}{24}\left(\frac{1}{L^d}\sum_{\vec{q}\neq 0}\frac{1}{\vec{q}^2+m_{2}^2}+\frac{1}{L^d}\sum_{\vec{q}\neq 0}\frac{(n-1)}{\vec{q}^2+m_{1}^2}\right)^2+\frac{g}{12}\left(\frac{1}{L^d}\sum_{\vec{q}\neq 0}\frac{1}{\vec{q}^2+m_{2}^2}\right)^2+\frac{g}{12}(n-1)\left(\frac{1}{L^d}\sum_{\vec{q}\neq 0}\frac{1}{\vec{q}^2+m_{1}^2}\right)^2\nonumber\\
 & -\frac{g^2 s^2}{12}\frac{1}{L^{2d}}\sum_{\substack{\{\vec{p},\vec{q}\}\neq 0,\\ \vec{p}\neq -\vec{q}}}\frac{1}{(\vec{p}^2+m_{2}^2)(\vec{q}^2+m_{2}^2)((\vec{p}+\vec{q})^2+m_{2}^2)}-\frac{g^2 s^2}{36}\frac{1}{L^{2d}}\sum_{\substack{\{\vec{p},\vec{q}\}\neq 0,\\ \vec{p}\neq -\vec{q}}}\frac{(n-1)}{(\vec{p}^2+m_{1}^2)(\vec{q}^2+m_{1}^2)((\vec{p}+\vec{q})^2+m_{2}^2)}\biggr)\nonumber\\
 &-\log{[\mathcal{N}']},
  \end{align}   
  with $m_{1}^2 = Z_{2}t+Z_{4}{g s^2}/{6}$, $m_{2}^2 = Z_{2}t+Z_{4}{g s^2}/{2}$ and  $\log{\left[\mathcal{N}\right]}-\log{\left[\mathcal{N}'\right]}=\mathcal{H}[\phi]|_{\phi=0}+\frac{1}{2}\Tr\log{\mathcal{H}^{(2)}_{M,s}}[\phi]|_{\phi=0}$.
\end{widetext}
In the MS scheme, the counter-terms $Z_{2}$, $Z_{4}$ and $\log{[\mathcal{N}']}$ at two loops are given by \cite{kleinert2001critical}:
\begin{equation}\label{counterterms}
\begin{split}
        Z_{2} & = 1 + \frac{n+2}{3\epsilon}\frac{\tilde{g}}{16\pi^2\epsilon}\\
    & +\frac{\tilde{g}^{2}}{(16\pi^2)^2}\left(\left(\frac{n+2}{3}\right)^2\frac{1}{\epsilon^2}+\frac{n+2}{3}\left(\frac{1}{\epsilon^2}-\frac{1}{2\epsilon} \right)\right),\\     
        Z_{4} & = 1 +  \frac{n+8}{3\epsilon}\frac{\tilde{g}}{16\pi^2}\\
    & +\frac{\tilde{g}^{2}}{(16\pi^2)^2}\left(\frac{3}{\epsilon^2}\frac{n^2+6n+20}{27}+\left(\frac{6}{\epsilon^2}-\frac{3}{\epsilon}\right)\frac{5n+22}{27}\right),\\
    \log{\left[\mathcal{N}'\right]} & = \tilde{t}^{2}\left(\frac{ n}{2(16\pi^{2})\epsilon}+\frac{\tilde{g}(n^2+2n)}{24(8\pi^2)^2\epsilon^{2}}\right) .
     \end{split}
    \end{equation}

The non-local divergences that appear as a result of the renormalization procedure should cancel out order by order in perturbation theory, thus leaving us with the same UV divergences in both the finite and the infinite volume. This has been explicitly shown in I for the Ising case and is also the case here. The computation of the diagrams appearing at two loops has been shown in the Appendix \ref{appenA} of this paper.

The quantity $\Gamma_M[\phi_{i}(x)=s_{i}]$ \textit{ a priori} depends upon a momentum scale $\mu$. At this scale $\mu$, $\Gamma_M[\phi_{i}(x)=s_{i}]$ can be equivalently parameterized in terms of dimensionless parameters defined at the same scale in the following way:
\begin{equation} \label{dimensionless}
\begin{split}
\bar{g}=&\mu^{-\epsilon}g =\mu^{-\epsilon}16\pi^2 \bar{u},\\
 \bar{s}=&\mu^{-1+\epsilon/2}Z^{-1/2}_{1}s,\\
\bar{t}=&\mu^{-2}t,\\
\bar{L}=&\mu L,
\end{split}
\end{equation}
where $Z_{1}=Z_{1}(\bar{u})$ is the field renormalization. Since we are working in a box of size $L$, it is natural to choose a scale $\mu=L^{-1}$ so as to include all the fluctuations between the UV scale $\Lambda$ and the infrared scale $\mu$ as fluctuations beyond this scale  are anyway forbidden. At this scale, the dimensionless variables are defined the following way (just as in \cite{sahu2024generalization, Sahu:2025bkp}):
\begin{equation}\label{stilde}
    \tilde{s}=sL^{\frac{d-2+\eta}{2}}\ , \ \tilde{g}=L^{4-d}{g}(\mu=L^{-1})\ , \tilde{t} = L^2 t(\mu = L^{-1}).   
\end{equation}\\
Universality takes place in the simultaneous limit of infinite volume $L\rightarrow\infty$ and criticality $t\rightarrow 0$. This renormalized temperature (difference) $t$  is linked to the bulk correlation length of the system as $\xi_{\infty} \sim t^{-1/\nu}$ asymptotically close to criticality. As shown in the one-loop calculation \cite{sahu2024generalization} and in \cite{balog2022critical, Rancon:2025bjf}, the double limit of criticality and infinite volume is not unique and may be approached in various ways keeping $\zeta=\lim_{L,\xi_{\infty}\rightarrow\infty}\frac{L}{\xi_{\infty}}$ fixed. Thus, we end up with a family of rate functions:
\begin{equation}
    \lim_{M\rightarrow\infty}\Gamma_{M}[\phi_{i}(x)=s_{i}] =L^{d}\hat{I}(s,\, \xi_{\infty},\,  L)= I_{\zeta, n}(\tilde{s})
\end{equation}
indexed by $\zeta$ (and also by $n$). Near criticality and at the scale $L^{-1}$ $\to 0$, we can directly replace $\tilde{g}$ by its fixed point value given by :
$\tilde{g}=16\pi^2 u^{(n)}_{*}$ with $u^{(n)}_{*}=\frac{3\epsilon}{n+8}+\frac{9\epsilon^2}{(n+8)^3}(3n+14)$ and the renormalized dimensionless temperature difference $\tilde{t}$ with $\zeta^{1/\nu}$.  

The rate function itself can be seen as the infinite volume fixed point potential with finite size corrections:
\begin{equation}\label{V,IV}
    I_{\zeta, n}(\tilde{s})=  I_{\zeta, n, {\rm inf}}(\tilde{s})+ I_{\zeta, n, {\rm fin}}(\tilde{s})
\end{equation}
where $ I_{\zeta, n, {\rm inf}}(\tilde{s})$ is the infinite volume fixed point potential and $I_{\zeta, n, {\rm fin}}(\tilde{s})$ encodes the finite size corrections. We show how to compute all the terms appearing in Eq.~\eqref{e1} in Appendix \ref{appenA}. Using  Eqs.~\eqref{e1},\eqref{counterterms} and defining the variable $x=\sqrt{u^{(n)}_{*}}\tilde{s}$, we now individually give the functions  $ I_{\zeta, n, {\rm fin}}(\tilde{s})$ and $I_{\zeta, n, {\rm inf}}(\tilde{s})$ up to second order in $\epsilon$:
\begin{widetext}
\begin{align}\label{ratef v}
      I_{\zeta, n, {\rm fin}}(x) & = \frac{1}{u_{*}}\biggl[\epsilon\biggl\{\frac{3}{n+8}\left(\frac{1}{2} (n-1)\Delta \left(\frac{\zeta ^{\frac{1}{\nu }}+8\pi^2{x^2}/{3}}{4\pi}\right)+\frac{1}{2} \Delta \left(\frac{\zeta ^{\frac{1}{\nu }}+8\pi^2  x^2}{4\pi}\right)\right)\biggr\}+\epsilon^2\bigg\{16\frac{\pi^2}{24}\left(\frac{3}{n+8}\right)^2\biggl((n^2-1)\biggl(\frac{1}{16\pi^2}\biggl\{\theta^{(1)}\left(\frac{\zeta ^{\frac{1}{\nu }}+\frac{8\pi^2}{3}  x^2}{4\pi}\right)\biggr\}^2\nonumber\\
      &+\frac{2}{4\pi}\theta^{(1)}\left(\frac{\zeta ^{\frac{1}{\nu }}+\frac{8\pi^2}{3}  x^2}{4\pi}\right)\left(\frac{\zeta ^{\frac{1}{\nu }}+\frac{8\pi^2}{3}  x^2}{(4\pi)^2}\right)\left(\gamma_{E}-1+\log{\left(\frac{\zeta ^{\frac{1}{\nu }}+\frac{8\pi^2}{3}  x^2}{4\pi}\right)}\right)\biggr)+3\biggl(\frac{1}{16\pi^2}\biggl\{\theta^{(1)}\left(\frac{\zeta ^{\frac{1}{\nu }}+8\pi^2 x^2}{4\pi}\right)\biggr\}^2\nonumber\\
      &+ \frac{2}{4\pi}\theta^{(1)}\left(\frac{\zeta ^{\frac{1}{\nu }}+8\pi^2  x^2}{4\pi}\right)\left(\frac{\zeta ^{\frac{1}{\nu }}+8\pi^2  x^2}{(4\pi)^2}\right)\left(\gamma_{E}-1+\log{\left(\frac{\zeta ^{\frac{1}{\nu }}+8\pi^2 x^2}{4\pi}\right)}\right)\biggr)+2(n-1)\biggl(\frac{1}{16\pi^2}\theta^{(1)}\left(\frac{\zeta ^{\frac{1}{\nu }}+\frac{8\pi^2}{3}  x^2}{4\pi}\right)\theta^{(1)}\left(\frac{\zeta ^{\frac{1}{\nu }}+8\pi^2  x^2}{4\pi}\right)\nonumber\\
      &+\frac{1}{4\pi}\theta^{(1)}\left(\frac{\zeta ^{\frac{1}{\nu }}+8\pi^2  x^2}{4\pi}\right)\left(\frac{\zeta ^{\frac{1}{\nu }}+\frac{8\pi^2}{3}  x^2}{(4\pi)^2}\right)\left(\gamma_{E}-1+\log{\left(\frac{\zeta ^{\frac{1}{\nu }}+\frac{8\pi^2}{3}  x^2}{4\pi}\right)}\right)+\frac{1}{4\pi}\theta^{(1)}\left(\frac{\zeta ^{\frac{1}{\nu }}+\frac{8\pi^2}{3}  x^2}{4\pi}\right)\left(\frac{\zeta ^{\frac{1}{\nu }}+8\pi^2  x^2}{(4\pi)^2}\right)\left(\gamma_{E}-1+\log{\left(\frac{\zeta ^{\frac{1}{\nu }}+8\pi^2 x^2}{4\pi}\right)}\right)\biggr)\biggr)\nonumber\\
      &-\frac{(16\pi^2)^2}{36}\left(\frac{3}{(n+8)}\right)^2 x^2\biggl(3\biggl(I_{1}\left[\zeta^{1/\nu}+8\pi^{2}{x}^2\right]+I_{2}\left[\zeta^{1/\nu}+8\pi^{2}{x}^2\right]+\frac{3}{16\pi^{2}\left(\zeta^{1/\nu}+8\pi^{2}{x}^2\right)}\theta^{(2)}\left(\frac{\zeta^{1/\nu}+8\pi^{2}{x}^2}{4\pi}\right)-\frac{1}{(\zeta^{1/\nu}+8\pi^{2}x^{2})^{3}}\nonumber\\
        & +\frac{3}{16\pi^2}\left(\log{4\pi}-\gamma_{E}-\log{\left(\zeta^{1/\nu}+8\pi^{2}{x}^2\right)}\right)\frac{1}{4\pi}\theta^{(1)}\left(\frac{\zeta^{1/\nu}+8\pi^{2}{x}^2}{4\pi}\right)\biggr)+(n-1)\biggl(2 \biggl(K_{1}\left[\zeta^{1/\nu}+8\pi^2 x^2, \zeta^{1/\nu}+\frac{8\pi^2}{3} x^2\right]\nonumber\\
       & +\frac{1}{(16\pi^2)}\left(\log{(4\pi)}-\gamma_{E}-\log{\left(\zeta^{1/\nu}+8\pi^2 x^2 \right)}\right)\left(\frac{1}{4\pi}\theta^{(1)}\left(\frac{\zeta^{1/\nu}+\frac{8}{3}\pi^2 x^2}{4\pi}\right)+\frac{1}{\zeta^{1/\nu}+\frac{8}{3}\pi^2 x^2}\right)\biggr)\nonumber\\
       &+K_{2}\left[\zeta^{1/\nu}+8\pi^2 x^2, \zeta^{1/\nu}+\frac{8\pi^2}{3} x^2\right]+\frac{1}{(16\pi^2)}\left(\log{(4\pi)}-\gamma_{E}-\log{\left(\zeta^{1/\nu}+\frac{8\pi^2}{3} x^2 \right)}\right)\left(\frac{1}{4\pi}\theta^{(1)}\left(\frac{\zeta^{1/\nu}+8\pi^2 x^2}{4\pi}\right)+\frac{1}{\zeta^{1/\nu}+8\pi^2 x^2}\right)\nonumber\\
        &+K_{3}\left[\zeta^{1/\nu}+8\pi^2 x^2, \zeta^{1/\nu}+\frac{8\pi^2}{3} x^2\right]-\frac{2}{\left(\zeta^{1/\nu}+\frac{8\pi^2}{3} x^2\right)\left(\frac{16\pi^2}{3} x^2\right)}\frac{1}{4\pi}\left(\theta^{(1)}\left(\zeta^{1/\nu}+\frac{8\pi^2}{3} x^2\right)-\theta^{(1)}\left(\zeta^{1/\nu}+8\pi^2 x^2\right)\right)\nonumber\\
        &-\frac{2}{16\pi^2\left(\zeta^{1/\nu}+\frac{8\pi^2}{3} x^2\right)\left(\frac{16\pi^2}{3} x^2\right)}\left(\left(\zeta^{1/\nu}+\frac{8\pi^2}{3}x^2\right)\left(\gamma_{E}-1+\log{\left(\frac{\zeta^{1/\nu}+\frac{8\pi^2}{3} x^2}{4\pi}\right)}\right)-\left(\zeta^{1/\nu}+8\pi^2 x^2\right)\left(\gamma_{E}-1+\log{\left(\frac{\zeta^{1/\nu}+8\pi^2 x^2}{4\pi}\right)}\right)\right)\nonumber\\
        &-\frac{1}{16\pi^2\left(\zeta^{1/\nu}+8\pi^2 x^2\right)}\left(\log{\left(4\pi\right)}-\gamma_{E}-\log{\left(\zeta^{1/\nu}+\frac{8\pi^2}{3} x^2\right)}-\theta^{(2)}\left(\frac{\zeta^{1/\nu}+\frac{8\pi^2}{3} x^2}{4\pi}\right)\right)-\frac{1}{\left(\zeta^{1/\nu}+8\pi^2 x^2\right)\left(\zeta^{1/\nu}+\frac{8\pi^2}{3} x^2\right)^2}\biggr)\biggr)\nonumber\\
        &-\frac{3}{n+8}\left(\frac{1}{2}\Delta^{(\epsilon)}\left(\frac{\zeta^{1/\nu}+8\pi^2 x^2}{4\pi}\right)+\frac{(n-1)}{2}\Delta^{(\epsilon)}\left(\frac{\zeta^{1/\nu}+\frac{8\pi^2}{3} x^2}{4\pi}\right)\right)+9\frac{3n+14}{(n+8)^3}\left(\frac{1}{2}\Delta\left(\frac{\zeta^{1/\nu}+8\pi^2 x^2}{4\pi}\right)+\frac{(n-1)}{2}\Delta\left(\frac{\zeta^{1/\nu}+\frac{8\pi^2}{3} x^2}{4\pi}\right)\right)\biggr\}\biggr],
\end{align}
Here $\gamma_{E}$ is the Euler-Mascheroni constant and the functions $\Delta$, $\Delta^{(\epsilon)}$, $\theta^{(1)}$, $\theta^{(2)}$, $I_{1}$, $I_{2}$, $K_{1}$, $K_{2}$ and $K_{3}$ are defined in the Appendix \ref{appenB} of the article.
\end{widetext}
\begin{widetext}
 \begin{align}\label{ratef iv}
        I_{\zeta, n, {\rm inf}}(x) & = \frac{1}{u_{*}}\Biggr[\frac{1}{2} \zeta ^{\frac{1}{\nu }} x^2 +\frac{2 \pi ^2 x^4}{3}+\epsilon\biggl\{\frac{3 }{{128 \pi ^2 (n+8)}}\biggl( (n-1)\left(\zeta ^{\frac{1}{\nu }}+\frac{8 \pi ^2 x^2}{3}\right)^2
   \left(2 \log \biggl(\frac{\zeta ^{\frac{1}{\nu }}+\frac{8 \pi ^2 x^2}{3}}{4\pi}\biggr)+2 \gamma_{E} -3\biggr)\right)\nonumber\\
   &+ \left(\zeta ^{\frac{1}{\nu }}+8 \pi ^2 x^2\right)^2 \left(2 \log \left(\frac{\zeta ^{\frac{1}{\nu }}+8 \pi ^2 x^2}{4 \pi }\right)+2 \gamma_{E}
   -3\right)\biggr)\biggr\}\nonumber\\
   & +\epsilon^2\Biggl\{\frac{1}{2}(n-1)\biggl(\frac{9(3n+14)}{64(n+8)^3\pi^2}\left(\zeta^{1/\nu}+\frac{8\pi^2 x^2}{3}\right)^2\biggl(2\log \biggl(\frac{\zeta ^{\frac{1}{\nu }}+\frac{8 \pi ^2 x^2}{3}}{4\pi}\biggr)+2\gamma_{E}-3\biggr)-\frac{1}{256(n+8)\pi^2}\left(\zeta^{1/\nu}+\frac{8\pi^2 x^2}{3}\right)^2\biggl(21+\pi^2\nonumber\\
   &+6 \gamma_{E} ^2-18 \gamma_{E} +6 (2 \gamma_{E} -3) \log \left(\frac{\zeta ^{\frac{1}{\nu }}+\frac{8 \pi ^2 x^2}{3}}{4 \pi }\right)+6 \log ^2\left(\frac{\zeta ^{\frac{1}{\nu }}+\frac{8 \pi ^2 x^2}{3}}{4 \pi }\right)\biggr)\biggr)+\frac{9(3n+14)}{128(n+8)^3\pi^2}\left(\zeta^{1/\nu}+8\pi^2 x^2\right)^2\biggl(2 \log \left(\frac{\zeta ^{\frac{1}{\nu }}+8 \pi ^2 x^2}{4 \pi }\right)\nonumber\\
   & +2 \gamma_{E}-3\biggr)-\frac{3}{4(n+8)^2}x^2\left(\zeta^{1/\nu}+8\pi^2 x^2\right)\left(-\frac{21}{2}-3A+9\gamma_{E}-3\gamma_{E}^2-\frac{\pi^2}{4}-(6\gamma_{E}-9)\log{\left(\frac{\zeta^{1/\nu}+8\pi^2 x^2}{4\pi}\right)}-3\log^{2}{\left(\frac{\zeta^{1/\nu}+8\pi^2 x^2}{4\pi}\right)}\right)\nonumber\\
   &-\frac{9}{128(n+8)^2\pi^2}\left(\zeta^{1/\nu}+8\pi^2 x^2\right)\left(\zeta^{1/\nu}+24\pi^2 x^2\right)\left(2-2\gamma_{E}+\gamma_{E}^2+\frac{\pi^2}{6}+2(\gamma_{E}-1)\log \left(\frac{\zeta ^{\frac{1}{\nu }}+8 \pi ^2 x^2}{4 \pi }\right)+\log^{2} \left(\frac{\zeta ^{\frac{1}{\nu }}+8 \pi ^2 x^2}{4 \pi }\right)\right)\nonumber\\
   &+\frac{9}{128(n+8)^2\pi^2}\left(\zeta^{1/\nu}+8\pi^2 x^2\right)^2\left(3-4\gamma_{E}+2\gamma_{E}^2+\frac{\pi^2}{6}+4(\gamma_{E}-1)\log \left(\frac{\zeta ^{\frac{1}{\nu }}+8 \pi ^2 x^2}{4 \pi }\right)+2\log^{2} \left(\frac{\zeta ^{\frac{1}{\nu }}+8 \pi ^2 x^2}{4 \pi }\right)\right)\nonumber\\
   &-\frac{1}{512(n+8)\pi^2}\left(\zeta^{1/\nu}+8\pi^2 x^2\right)^2\biggl(21+\pi^2
   +6 \gamma_{E} ^2-18 \gamma_{E} +6 (2 \gamma_{E} -3) \log \left(\frac{\zeta ^{\frac{1}{\nu }}+8 \pi ^2 x^2}{4 \pi }\right)+6 \log ^2\left(\frac{\zeta ^{\frac{1}{\nu }}+8 \pi ^2 x^2}{4 \pi }\right)\biggr)\nonumber\\
   &+\frac{1}{768(n+8)^2\pi^2}(n-1)\Biggl(-\left(\zeta^{1/\nu}+\frac{8\pi^2 x^2}{3}\right)\left(3(n+2)\zeta^{1/\nu}+8(n+8)\pi^2 x^2\right)\biggl(12+6\gamma_{E}^2-12\gamma_{E}+\pi^2+12(\gamma_{E}-1)\log\left(\frac{\zeta^{1/\nu}+\frac{8\pi^2 x^2}{3}}{4\pi}\right)\nonumber\\
   &+6\log^{2}\left(\frac{\zeta^{1/\nu}+\frac{8\pi^2 x^2}{3}}{4\pi}\right)\biggr)-3\left(\zeta^{1/\nu}+8\pi^2 x^2\right)^2\biggl(12+6\gamma_{E}^2-12\gamma_{E}+\pi^2+12(\gamma_{E}-1)\log\left(\frac{\zeta^{1/\nu}+8\pi^2 x^2}{4\pi}\right)+6\log^{2}\left(\frac{\zeta^{1/\nu}+8\pi^2 x^2}{4\pi}\right)\biggr)\nonumber\\
   &+3\left(\zeta^{1/\nu}+\frac{8\pi^2}{3} x^2\right)\biggl((n+1)\left(\zeta^{1/\nu}+\frac{8\pi^2}{3} x^2\right)\left(18-24\gamma_{E}+12\gamma_{E}^2+\pi^2+24(\gamma_{E}-1)\log\left(\frac{\zeta^{1/\nu}+\frac{8\pi^2}{3} x^2}{4 \pi}\right)+12\log^{2}\left(\frac{\zeta^{1/\nu}+\frac{8\pi^2}{3} x^2}{4 \pi}\right)\right)\nonumber\\
   &+2\left(\zeta^{1/\nu}+8\pi^2 x^2\right)\biggl(18-24\gamma_{E}+12\gamma_{E}^2+\pi^2+3\log^{2}\left(\frac{\zeta^{1/\nu}+\frac{8\pi^2}{3} x^2}{4\pi}\right)+12(\gamma_{E}-1)\log\left(\frac{\zeta^{1/\nu}+\frac{8\pi^2}{3} x^2}{4\pi}\right)+3\log^{2}\left(\frac{\zeta^{1/\nu}+8\pi^2 x^2}{4\pi}\right)\nonumber\\
   &+12(\gamma_{E}-1)\log\left(\frac{\zeta^{1/\nu}+8\pi^2 x^2}{4\pi}\right)+6\log\left(\frac{\zeta^{1/\nu}+8\pi^2 x^2}{4\pi}\right)\log\left(\frac{\zeta^{1/\nu}+\frac{8\pi^2}{3} x^2}{4\pi}\right)\biggr)\biggr)+\frac{8}{3}\pi^2 x^2\biggl(18 \pi ^2 \zeta ^{\frac{1}{\nu }}+216 \gamma_{E} ^2 \zeta ^{\frac{1}{\nu }}-648 \gamma_{E}  \zeta
   ^{\frac{1}{\nu }}\nonumber\\
   &+756 \zeta ^{\frac{1}{\nu }}+80 \pi ^4 x^2+960 \gamma_{E} ^2 \pi ^2 x^2-2880 \gamma_{E}  \pi ^2 x^2+3360 \pi ^2 x^2-192 \pi ^2 x^2 \sqrt{\frac{54 \zeta ^{\frac{1}{\nu }}+48 \pi ^2 x^2}{2 \zeta ^{\frac{1}{\nu }}+16 \pi ^2 x^2}} B\left[\zeta ^{\frac{1}{\nu }}+8 \pi ^2
   x^2,\zeta ^{\frac{1}{\nu }}+\frac{8 \pi ^2 x^2}{3}\right]\nonumber\\
   &+24 \zeta ^{\frac{1}{\nu }} \sqrt{\frac{54 \zeta ^{\frac{1}{\nu }}+48 \pi ^2 x^2}{2 \zeta ^{\frac{1}{\nu }}+16 \pi ^2 x^2}} B\left[\zeta ^{\frac{1}{\nu
   }}+8 \pi ^2 x^2,\zeta ^{\frac{1}{\nu }}+\frac{8 \pi ^2 x^2}{3}\right]+216 \zeta ^{\frac{1}{\nu }}\log ^2{4 \pi } -432 \gamma_{E} \zeta
   ^{\frac{1}{\nu }} \log {4 \pi } +648\zeta ^{\frac{1}{\nu }} \log {4 \pi} +960 \pi ^2 x^2 \log ^2{4 \pi }\nonumber\\
   &-1920 \gamma_{E}  \pi ^2 x^2 \log {4 \pi  }+2880 \pi ^2 x^2 \log{4 \pi }+12 \left(18 \zeta ^{\frac{1}{\nu }}+80 \pi ^2 x^2\right) \log ^2\left(\zeta ^{\frac{1}{\nu }}+8 \pi ^2 x^2\right)+6 \left(18 \zeta ^{\frac{1}{\nu }}+16
   \pi ^2 x^2\right) \log ^2\left(\frac{6 \zeta ^{\frac{1}{\nu }}+16 \pi ^2 x^2}{6 \zeta ^{\frac{1}{\nu }}+48 \pi ^2 x^2}\right)\nonumber\\
   &+24 \left(-3+2 \gamma_{E} -2 \log{4 \pi }\right) \left(6 \zeta ^{\frac{1}{\nu }}+16 \pi ^2 x^2\right) \log \left(\frac{6 \zeta ^{\frac{1}{\nu }}+16 \pi ^2 x^2}{6 \zeta ^{\frac{1}{\nu }}+48
   \pi ^2 x^2}\right)+12 \log \left(\zeta ^{\frac{1}{\nu }}+8 \pi ^2 x^2\right) \biggl(\left(-3+2 \gamma_{E} -2 \log{4 \pi}\right)\left(18 \zeta ^{\frac{1}{\nu }}+80 \pi ^2 x^2\right)\nonumber\\
   &+4\left(6 \zeta ^{\frac{1}{\nu }}+16 \pi ^2 x^2\right) \log \left(\frac{6 \zeta ^{\frac{1}{\nu }}+16 \pi ^2 x^2}{6 \zeta ^{\frac{1}{\nu }}+48 \pi ^2
   x^2}\right)\biggr)\biggr)\Biggr)\Biggr\}\Biggl].
   \end{align}
   Here the constant $A$ and the function $B$ are defined in the appendix \ref{appenB} of this article.
\end{widetext}
We stress here the fact that using the variable $x$ rather than $\tilde{s}$ is on one hand a finite redefinition of the field when $\epsilon$ is fixed and non-vanishing and on the other hand guarantees a systematic $\epsilon$-expansion. We also note that the $\epsilon\rightarrow 0$ limit of the theory is ill-defined and hence the limit of $d=4$ cannot be studied using the current framework. However this is not really a surprise as explained in \cite{lawrie1976field}. \\

The derivation above has been done, by defining the renormalized parameters at the scale $\mu=L^{-1}$. We could have done the derivation by defining the renormalized parameters at any scale $\mu$. This entails in demanding that the rate function is a scaling function of the variables $\tilde{s}$ and $\zeta$. This has already been shown at one loop\cite{sahu2024generalization, eisenriegler1987helmholtz} and at two loops for the Ising model in I.\\

As shown in Eq.~\eqref{V,IV}, the rate function can be interpreted as the effective potential with finite size corrections. This implies that the rate function inherits the large field behavior of the effective potential. This is indeed the case as one can easily show:
\begin{align}\label{LF1}
    I_{\zeta, n, {\rm inf}}(x)\underset{x\to\infty}{\sim}
        & \frac{x^{4}}{u^{(n)}_{*}}  \biggl(1  +\epsilon\log{x}+\frac{\epsilon^{2}}{2(n+8)^2}((n^2+14n+60)\log{x}\nonumber\\
    &+(n+8)^2\log^{2}x)\biggr),
\end{align}
which is the expansion of $x^{\delta+1}$ for $O(n)$ models up to second order in $\epsilon$ with $\delta=3+\epsilon+\frac{n^2+14n+60}{2(n+8)^2}\epsilon^2$. However there is also a sub-leading universal large field behavior of the rate function due to the finite size volume corrections. Thus in the limit $x\rightarrow\infty$, we have:
\begin{widetext}
\begin{align}\label{LF2}
   I_{\zeta, n, {\rm fin}}(x)\underset{x\rightarrow\infty}\sim & \frac{1}{u^{(n)}_{*}}\left(\frac{3\epsilon}{n+8}+\frac{9(3n+14)}{(n+8)^3}\epsilon^2\right)\left(\frac{1}{2}\Delta\left(x^2\right)+\frac{(n-1)}{2}\Delta\left(\frac{x^2}{3}\right)\right)+\frac{1}{u^{(n)}_{*}}\left(\frac{3}{n+8}\right)^2\epsilon^2\biggl[\frac{1}{24}\biggl(2(n^2-1)\frac{x^2}{3}\theta^{(1)}\left(\frac{x^2}{3}\right)\log{\left(\frac{x^2}{3}\right)}\nonumber\\
     &+6x^2\theta^{(1)}\left(x^2\right)\log{\left(x^2\right)}+2(n-1)\biggl(\frac{x^2}{3}\theta^{(1)}(x^2)\log{\left(\frac{x^2}{3}\right)}+x^2\theta^{(1)}\left(\frac{x^2}{3}\right)\log{\left(x^2\right)}\biggr)\biggr)+\frac{16}{18} x ^2\biggl(\frac{9}{16}~\theta^{(1)}(x^2)\log{(x^2)}\nonumber\\
     &+(n-1)\biggl(\frac{2}{16}~\theta^{(1)}\left(\frac{x^2}{3}\right)\log{(x^2)}+\frac{1}{16}\log{\left(\frac{x^2}{3}\right)}\theta^{(1)}\left(x^2\right)\biggr)\biggr)\biggr]\nonumber\\
     &\underset{x\rightarrow\infty}\sim -n\left(1+\frac{\epsilon}{2}+O(\epsilon^2)\right)\log{x}
\end{align}
where in the last line we have used the fact that in the limit $x\rightarrow\infty$ the functions $\Delta(x^{2})$ and  $x^{2}\theta^{(1)}(x^{2})\log{x^{2}}$ go as $ -\log{x^{2}}$ as shown in \cite{Sahu:2025bkp}. (Factors of $\pi$ have been suppressed in the calculation above for brevity reasons.)
\end{widetext}
This sub-leading correction given by Eq.~\eqref{LF2} is nothing but $n\frac{\delta-1}{2}$ expanded up to first order in $\epsilon$. Thus using Eqs.~\eqref{LF1} and \eqref{LF2} one can exactly predict analytically the large field behaviour of the rate function $I_{\zeta,n}(x)$ which is given by: $I_{\zeta,n}(x)\underset{x\rightarrow\infty}{\sim} x^{\delta+1}-n\frac{(\delta-1)}{2}\log{x}$. This behaviour has already been seen at one loop and has also been predicted for the Ising model some time ago in \cite{Hilfer1995, bruce1995critical}, can also be seen exactly in the large-$n$ limit \cite{balog2024universal, Rancon:2025bjf} and is in fact generic for equilibrium systems as shown in \cite{Stella2023, balog2024universal}.

\section{Comparison with Monte-Carlo Simulations}

As shown in I and in \cite{privman1984universal, bervillier1976universal}, for comparison of the rate function obtained from MC simulations and from perturbation theory one needs to \textit {a priori} get rid of two non-universal factors. Thus we have two scales to fix : the scale of the order parameter and the temperature scale. These scales are fixed in the following way: the scale of the order parameter is fixed by demanding that the minimum of the rate function obtained from MC simulations and from perturbation theory for $\zeta=0$ coincide, while the temperature scale is fixed by demanding that the curvature of the rate function obtained from MC simulations and from perturbation theory  i.e. $\frac{\partial I_{\zeta}(\rho)}{\partial \rho}|_{\rho=0}$ (with $\rho=\tilde{s}^{2}/2$) for some $\zeta\neq0$ coincide. Of course, for $\zeta=0$, the temperature scale need not be fixed. The same scale-fixing procedure has been used in I.

\begin{figure}[t]
\centering
\includegraphics[width=\linewidth]{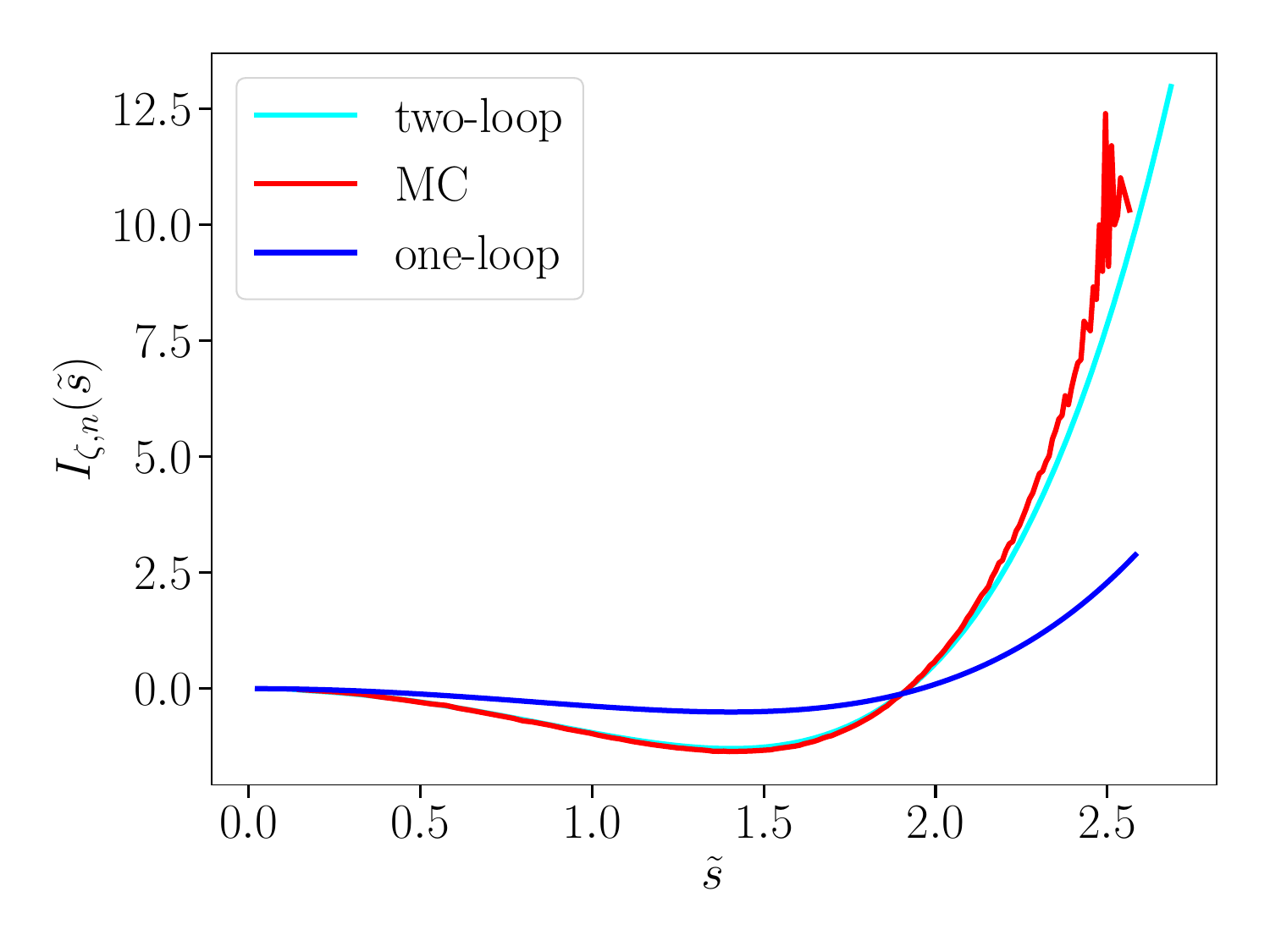}
\caption{ \label{Rate0O22L.pdf} Comparison of the rate function, $I_{\zeta=0,n=2}(x)$ obtained from Monte Carlo (MC) simulations (red) and from one-loop and two-loop order (cyan) of perturbation theory for the $O(2)$ model (Classical XY model). }
\end{figure}

\begin{figure}[t]
\centering
\includegraphics[width=\linewidth]{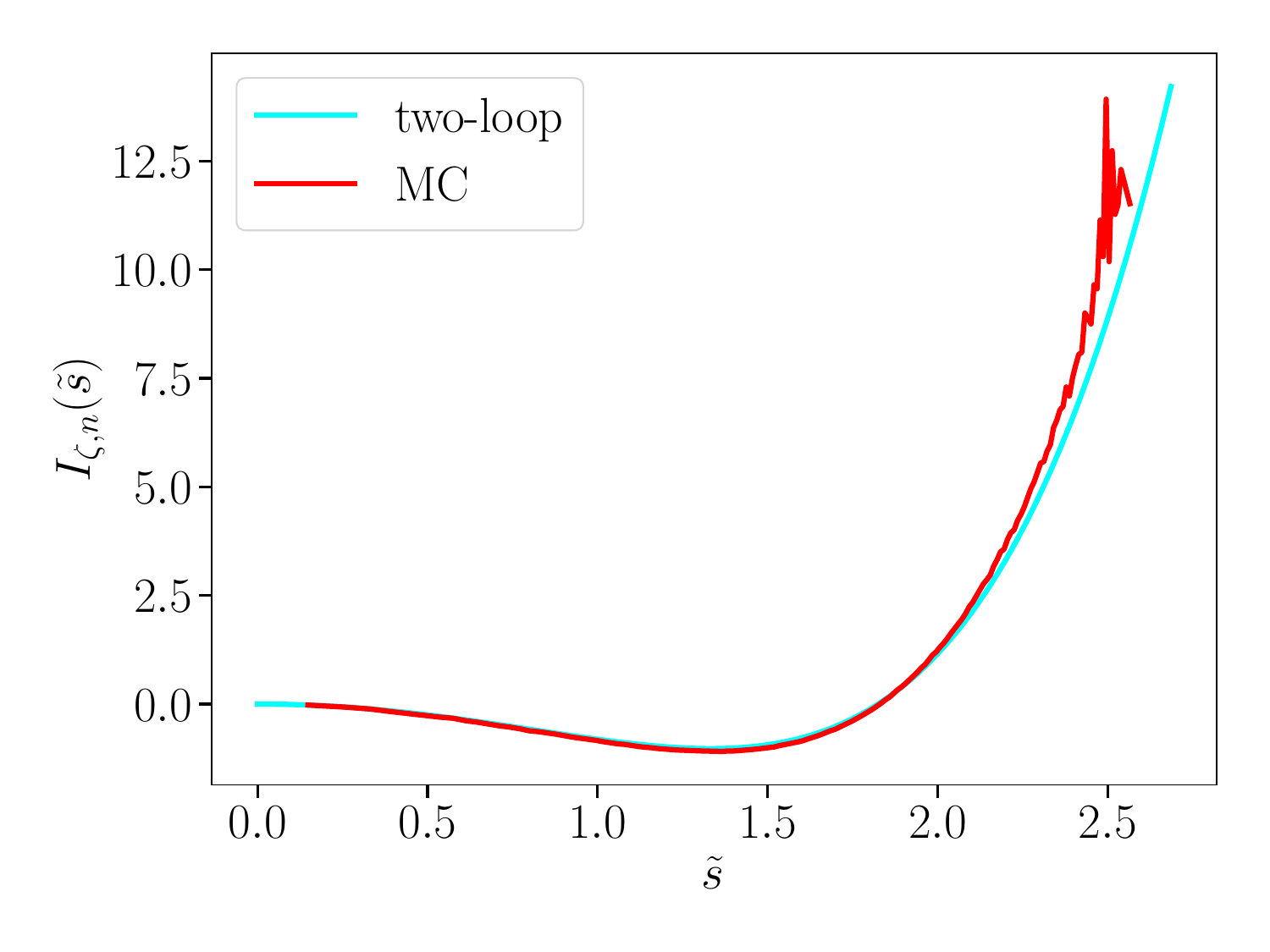}
\caption{ \label{Rate0.5O22L.pdf} Comparison of $I_{\zeta=0.5,n=2}(x)$ obtained from Monte Carlo (MC) simulations (red) and from two-loop order (cyan) of perturbation theory. }
\end{figure}

\begin{figure}[t]
\centering
\includegraphics[width=\linewidth]{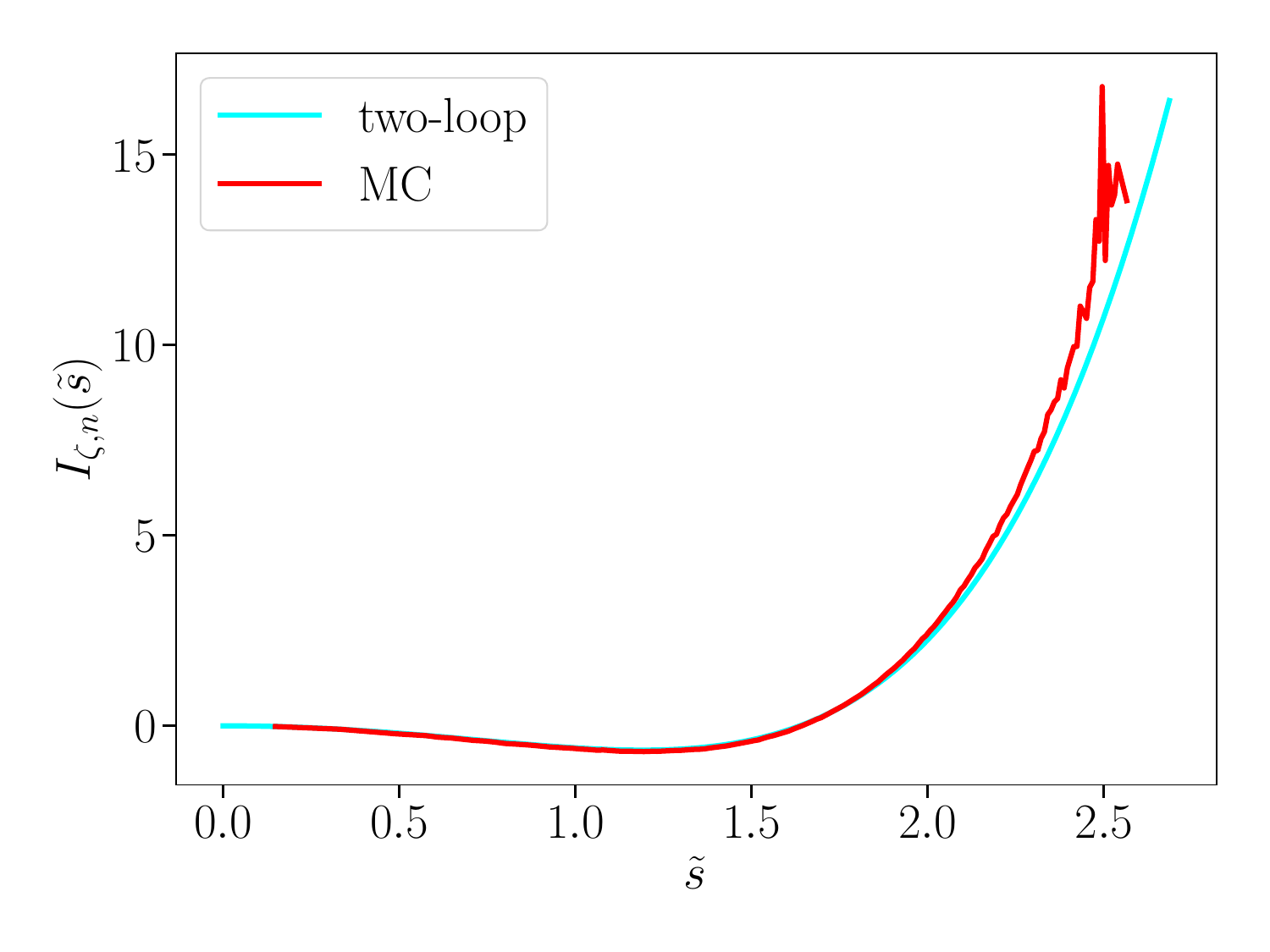}
\caption{ \label{Rat1O22L.pdf} Comparison of $I_{\zeta=1,n=2}(x)$ obtained from Monte Carlo (MC) simulations (red) and from two-loop order (cyan) of perturbation theory. }
\end{figure}

\begin{figure}[t]
\centering
\includegraphics[width=\linewidth]{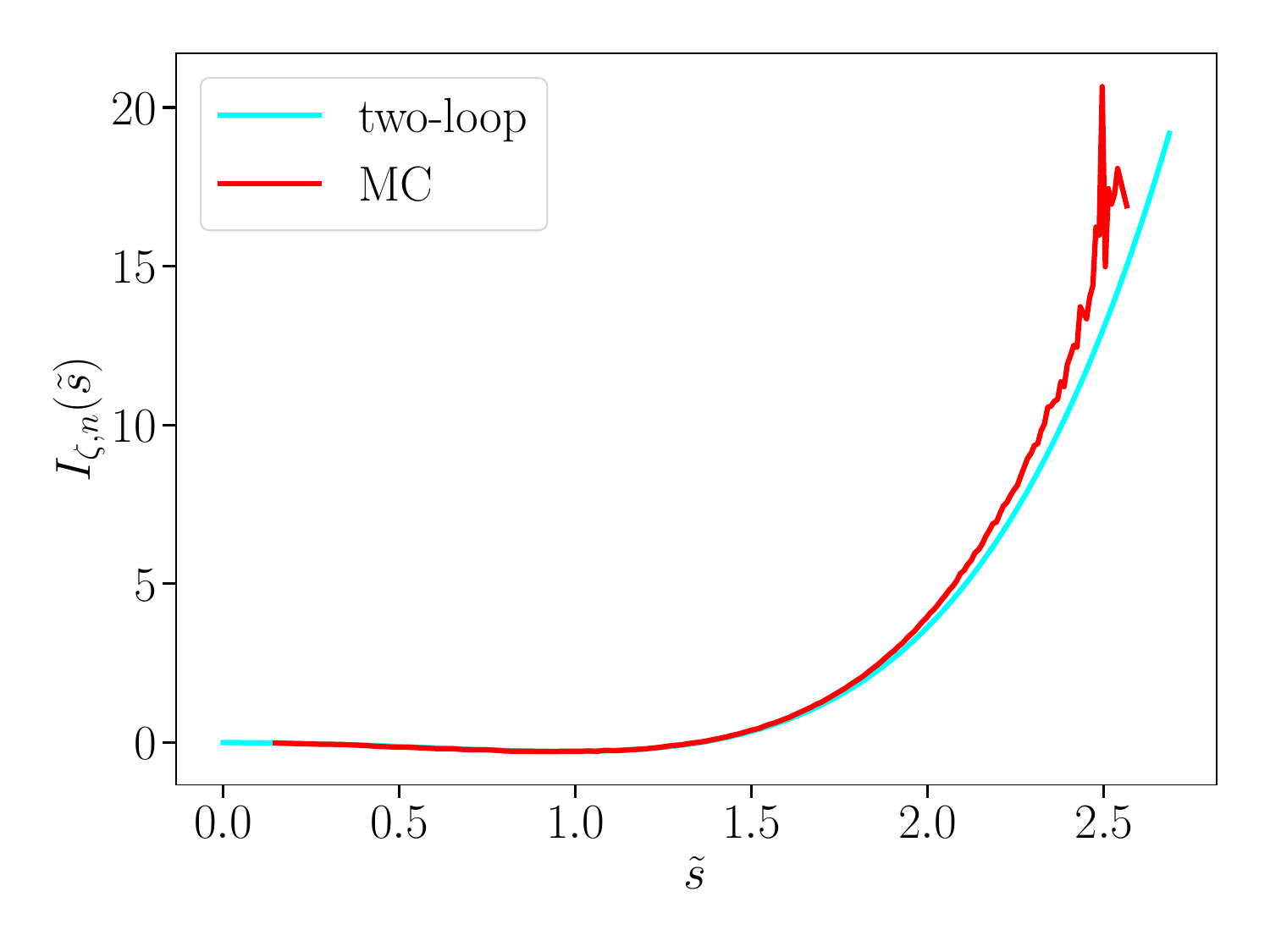}
\caption{ \label{Rat1.5O22L.pdf} Comparison of $I_{\zeta=1.5,n=2}(x)$ obtained from Monte Carlo (MC) simulations (red) and from two-loop order (cyan) of perturbation theory. }
\end{figure}

\begin{figure}[t]
\centering
\includegraphics[width=\linewidth]{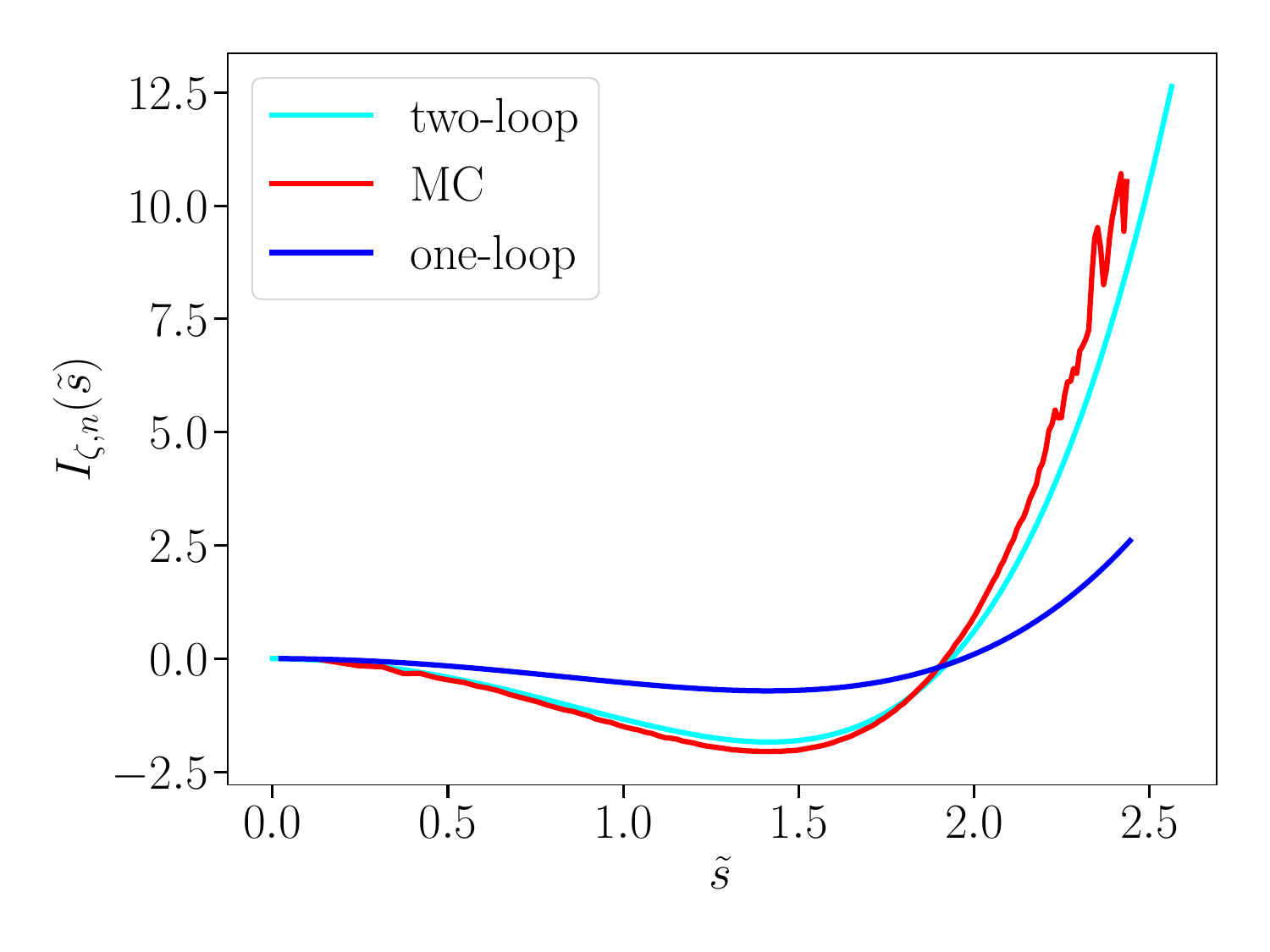}
\caption{ \label{Rate0O32L.pdf} Comparison of $I_{\zeta=0,n=3}(x)$ obtained from Monte Carlo (MC) simulations (red) and from one-loop (blue) and two-loop order (cyan) of perturbation theory for the $O(3)$ model (Heisenberg model). }
\end{figure}

\begin{figure}[t]
\centering
\includegraphics[width=\linewidth]{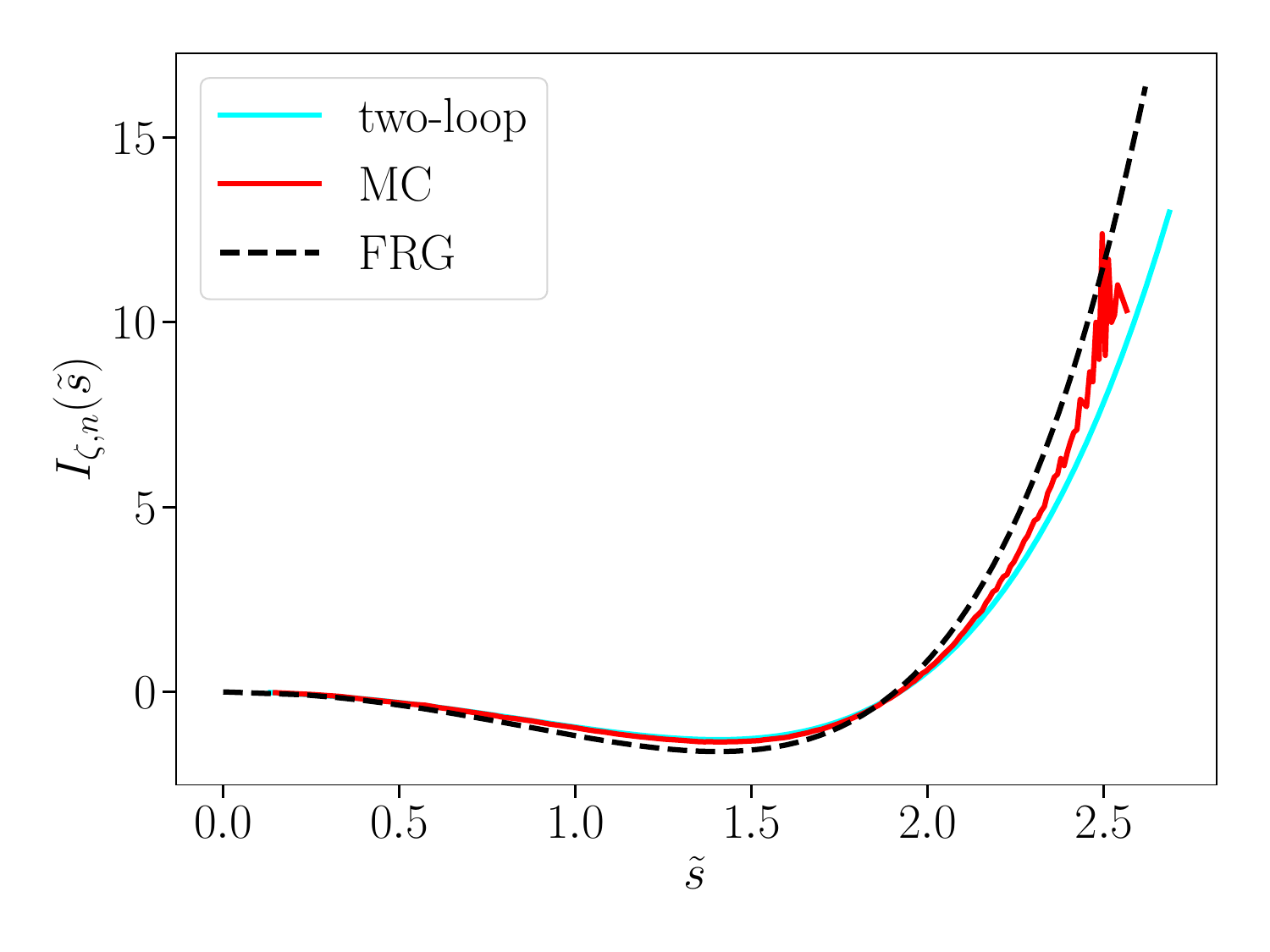}
\caption{ \label{Rate0O22LFRG.pdf} Comparison of $I_{\zeta=0,n=2}(x)$ obtained from Monte Carlo (MC) simulations (red), from two-loop order of perturbation theory (cyan) and from the FRG Calculations (LPA)\cite{Rancon:2025bjf} for the $O(2)$ model (Classical XY model). }
\end{figure}

\begin{figure}[t]
\centering
\includegraphics[width=\linewidth]{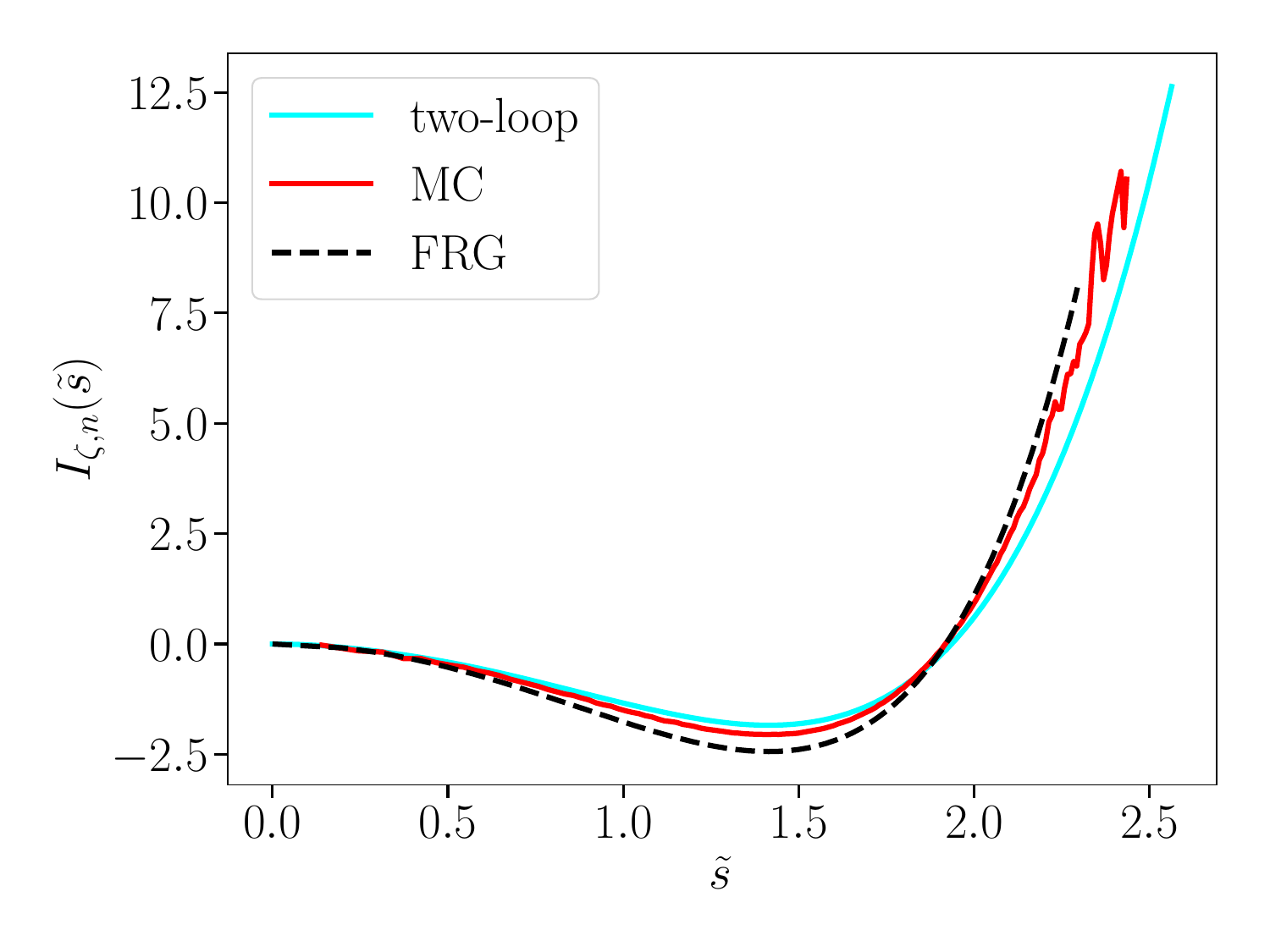}
\caption{ \label{Rate0O32LFRG.pdf} Comparison of $I_{\zeta=0,n=3}(x)$ obtained from Monte Carlo (MC) simulations (red), from two-loop order of perturbation theory (cyan) and from the FRG Calculations (LPA) \cite{Rancon:2025bjf} for the $O(3)$ model (Heisenberg model). }
\end{figure}

We find that after fixing these two scales the agreement of the MC data with the two-loop results are much better compared to its one-loop counterpart as shown in Figs. \ref{Rate0O22L.pdf} and \ref{Rate0O32L.pdf} for the $O(2)$ (classical XY spin model) and the $O(3)$ (Heisenberg model) model respectively (for $\zeta=0$). We further improve the determination of the minimum of the rate function at $\zeta=0$ for the $O(2)$ and $O(3)$ model as suggested by the Tables \ref{tabImin} and \ref{tabImin3}. In Figs. \ref{Rate0.5O22L.pdf}, \ref{Rat1O22L.pdf} and \ref{Rat1.5O22L.pdf}, we have further compared the rate function $I_{\zeta, n}(\tilde{s})$ obtained from the MC data\footnote{The tails of the rate function obtained from Monte-Carlo Simulations are wobbly because of the lack of statistics at the tails of the distribution.} and at two loops for higher values of $\zeta$ for the $O(2)$ model. We find that although the small field behavior is rather well reproduced at two loops, the same does not hold true for the large field behavior. This is however expected, since as shown in the previous section the large field behavior is dominated by a single exponent $\delta$, and what we retrieve at two loops is not the true exponent $\delta$ but rather its $\epsilon-$expansion up to second order in $\epsilon$. To cure the large field behavior, we could very well perform a RG improvement, however this is beyond the scope of the current article and is in fact rather demanding to perform.\\
We also compare our results with the existing FRG calculations \cite{Rancon:2025bjf} for the $O(2)$ and the $O(3)$ model for $\zeta=0$ as shown in Figs. \ref{Rate0O22LFRG.pdf} and \ref{Rate0O32LFRG.pdf} respectively.
\begin{table}[t]
  \begin{center}
    \begin{tabular}{ c|c } 
      \text{Methods} 
       &$I^{min}_{\zeta=0, n=2}$\\
      \hline
       \\
 \text{  Mean Field} & $0$ \\
     &\\
       \text{One-loop} \cite{Rancon:2025bjf} & $-0.503$\\
     &\\
     \text{Two-loop} & $-1.292$\\
     &\\
      \text{Monte-Carlo} & $-1.354$\\
     &\\
       \text{FRG} (LPA) \cite{Rancon:2025bjf} & $-1.614$\\
    \end{tabular}
    \caption{\label{tabImin} Determination of the minimum of the rate function $I_{\zeta=0, n=2}(x)$ i.e. $I^{\rm{min}}_{\zeta=0, n=2}$ via different methods.}
  \end{center}
\end{table}

\begin{table}[t]
  \begin{center}
    \begin{tabular}{ c|c } 
      \text{Methods} 
       &$I^{min}_{\zeta=0, n=3}$\\
      \hline
       \\
 \text{  Mean Field} & $0$ \\
     &\\
       \text{One-loop}\cite{Rancon:2025bjf} & $-0.712$\\
     &\\
     \text{Two-loop} & $-1.841$\\
     &\\
      \text{Monte-Carlo} & $-2.030$\\
     &\\
       \text{FRG} (LPA) \cite{Rancon:2025bjf} & $-2.437$\\
    \end{tabular}
    \caption{\label{tabImin3} Determination of the minimum of the rate function $I_{\zeta=0, n=3}(x)$ i.e. $I^{\rm{min}}_{\zeta=0, n=3}$ via different methods.}
  \end{center}
\end{table}

\section{Conclusions}
In this article, we show how to compute the critical PDFs or equivalently its their logarithm which is the rate function of the order parameter of the $O(n)$ model perturbatively at second order of the $\epsilon=4-d$ expansion. It has already been shown in \cite{balog2022critical, Rancon:2025bjf}, at one loop \cite{sahu2024generalization} and at two loops \cite{Sahu:2025bkp} (for Ising, $n=1$) that there exists not one but a family of PDFs at criticality indexed by $\zeta=L/\xi_\infty$ depending on the way the thermodynamic and critical limits are taken. We find that the PDFs computed at two loops compare very well with the Monte-Carlo simulations compared to its one-loop counterpart as shown in the previous section. However, although the small field region is rather well reproduced at two loops, the large field behavior is not as good which calls for RG improvement as shown in \cite{guida19973d, chung1999renormalization}. This is beyond the scope the current article but is in plans for our future work. However we must notice that in the Ising case, even at two loops the large field region is rather well reproduced contrary to what happens for $n>1$. The reason for this is not yet obvious and would require further examination. We stress here that these PDFs although universal have a dependence on boundary conditions. For the current article, we have only studied them in presence of periodic boundary conditions. Hence, a natural continuation of the present work would be to see the effect and dependence of other boundary conditions on the PDF. \\

In the non-perturbative (FRG) methods, one can rather easily compute these PDFs in the ordered phase or at $\zeta<0$ \cite{Rancon:2025bjf, balog2022critical}. While at one loop we were also able to generalize our results to  the low temperature phase i.e. $\zeta< 0$ rather easily via analytical continuation of the results obtained in the disordered phase, the same is not true at two loops and requires further investigation. One of the most common perturbative methods that is deployed for studying the $O(n)$ model is the $1/n$ expansion. Already these PDFs have been computed exactly for $O(n=\infty)$ case in \cite{Rancon:2025bjf, balog2024universal} which marks the zeroth order of the  $1/n$ expansion. Thus perturbatively, one could also consider studying these PDFs in the $1/n$ expansion. This is a problem we would like to tackle in the future.\\

As suggested in \cite{lawrie1976field} and pointed out in Section \ref{3} of the current work, our framework to study these critical PDFs of the order parameter breaks down right at $d=4$. We are  keen to develop a method that would help us to study these critical PDFs at $d=4$ and thereby study the problem of triviality in field theories in a more detailed way.\\ 

Although we have successfully developed a perturbative approach for computing the PDFs of the order parameter of equilibrium systems like the $O(n)$ and the Ising model, it will be interesting to see if we could generalize our framework for studying these PDFs in out-of-equilibrium systems like reaction-diffusion systems, directed percolation, etc.

\section{Acknowledgements}
The author thanks Bertrand Delamotte for his support during the work and for his valuable comments on the initial draft of this manuscript. The author also thanks Adam Ran\c con for his comments on the manuscript. The author also thanks Adam Ran\c con and Ivan Balog for many fruitful discussions and for providing him with the Monte-Carlo and the FRG(LPA) data. 

\newpage
\appendix
\onecolumngrid
\section{Computing the discrete sums}\label{appenA}
In this section, we show how to compute the discrete sums in Eq.~\eqref{e1}. In I \cite{Sahu:2025bkp}, we have already shown how to compute the homogenous sunset like term (the diagram appearing in $\mathcal{I}_{1}$). Here we use similar techniques to compute the inhomogenous sunset like term (the diagram appearing in $\mathcal{I}_{2}$).\\

We start by defining the function $\theta(z)$.
In \cite{Sahu:2025bkp}, it has already been shown:
\begin{equation}
    \frac{1}{L^d}\sum_{q\neq 0}\log{\left(1+\frac{r}{\vec{q}^2}\right)} = \frac{1}{L^d}\left(\theta\left(L^2r /4\pi\right)-\theta(0)\right)+\int \frac{d^{d}k}{(2\pi)^d}\log\left(1+\frac{r}{k^2}\right),
\end{equation}
with:
\begin{equation}
\theta(z)=-\int_0^\infty d\sigma \frac{e^{-\sigma z}}\sigma\left(\vartheta^d(\sigma)-1-\sigma^{-d/2}\right).
\end{equation}
where $\vartheta(\sigma)$ is the Jacobi-theta function given by: $\vartheta(\sigma)=\sum_{j=-\infty}^{\infty}e^{-j^2 \pi \sigma}$.\\
Derivating it $n$ times with respect to $r$ one hence obtains:
\begin{equation}\label{discrete}
    \frac{1}{L^d}\sum_{q\neq 0}\frac{1}{\left(q^2+r\right)^n} = \frac{(-1)^{n-1}}{(n-1)!}\frac{1}{L^d}\left(\frac{L^2}{4\pi}\right)^n\theta^{(n)}\left(\frac{L^2r}{4\pi}\right)+\int \frac{d^{d}k}{(2\pi)^d}\frac{1}{\left(k^2+r\right)^n},
\end{equation}
where $\theta^{(n)}\left(m\right)$ means taking the derivative of $\theta\left(m\right)$, $n$ times with respect to $m$. \\

The inhomogeneous sunset like term is given by:
\begin{align}\label{sunin}
& \frac{1}{L^{2d}}\left(\sum_{\substack{\{\vec{p},\vec{q}\}\neq 0,\\ \vec{p}\neq -\vec{q}}}\frac{1}{(\vec{p}^2+m_{1}^2)(\vec{q}^2+m_{1}^2)((\vec{p}+\vec{q})^2+m_{2}^2)}\right)\nonumber\\
& =\frac{1}{L^{2d}}\left(\sum_{\vec{p},\vec{q}}\frac{1}{(\vec{p}^2+m_{1}^2)(\vec{q}^2+m_{1}^2)((\vec{p}+\vec{q})^2+m_{2}^2)}-\frac{1}{m^2_{2}}\sum_{\vec{p}\neq 0}\frac{1}{(\vec{p}^2+m^2_{1})^2}-\frac{2}{m_{1}^2}\sum_{\vec{p}\neq 0}\frac{1}{(\vec{p}^2+m_{1}^2)(\vec{p}^2+m_{2}^2)}-\frac{1}{(m_{2}^2)^2 m^2_{1}}\right).
\end{align} 
Using the relation \eqref{discrete} one can easily compute the second and the third term in Eq.~\eqref{sunin}. These are given by:
\begin{align}
    \frac{1}{L^{2d} m^2_{1}}\sum_{\vec{p}, \vec{q}\neq 0}\frac{1}{(\vec{p}^2+m^{2}_{1})(\vec{p}^2+m^{2}_{2})} & = \frac{1}{L^{2d-6}}\biggl(\frac{1}{m^{2}_{1}L^2(m^{2}_{2}L^2-m^{2}_{1}L^2)}\biggl(\frac{1}{4\pi}\theta^{(1)}\left(\frac{m^2_{1} L^2}{4\pi}\right)-\frac{1}{4\pi}\theta^{(1)}\left(\frac{m^2_{2} L^2}{4\pi}\right)\biggr)\nonumber\\
    & + \frac{1}{m^{2}_{1}L^2(m^{2}_{2}L^2-m^{2}_{1}L^2)}\left(\frac{(m^{2}_{2}L^2-m^{2}_{1}L^2)}{8\pi^2\epsilon}+\frac{(m^2_{1}L^2)}{16\pi^2}\left(\gamma_{E}-1+\log{\frac{m^{2}_{1}L^2}{4\pi}}\right)-\frac{(m^2_{2}L^2)}{16\pi^2}\left(\gamma_{E}-1+\log{\frac{m^{2}_{2}L^2}{4\pi}}\right)\right)\biggr),
\end{align}
and
\begin{equation}
   \frac{1}{L^{2d}}\sum_{q\neq 0}\frac{1}{m_{2}^2\left(q^2+m_{1}^2\right)^2} = \frac{1}{L^{2d-6}}\left(\frac{1}{16\pi^2 (m_{2}^2 L^2)}\left(\frac{2}{\epsilon}+\log{4\pi}-\gamma_{E}-\log{(m_{1}^2 L^2)}\right)-\frac{1}{16\pi^2 (m_{2}^2L^2)}\theta^{(2)}\left(\frac{m_{1}^2 L^2}{4\pi}\right)\right) .
\end{equation}

We now show how to compute the first term in Eq.~\eqref{sunin}.\\
Using Poisson summation we have:
  \begin{align}\label{sunin2}
& \frac{1}{L^{2d}}\sum_{\vec{p},\vec{q}}\frac{1}{(\vec{p}^2+m_{1}^2)(\vec{q}^2+m_{1}^2)((\vec{p}+\vec{q})^2+m_{2}^2)}\nonumber\\
& = \int \frac{d^d{p}}{(2\pi)^d}~\frac{d^{d}{q}}{(2\pi)^d}\frac{1}{(\vec{p}^2+m_{1}^2)(\vec{q}^2+m_{1}^2)((\vec{p}+\vec{q})^2+m_{2}^2)}\nonumber\\
& +2 \int \frac{d^d{p}}{(2\pi)^d}~\frac{d^{d}{q}}{(2\pi)^d}\sum_{l_{p}\neq 0}\frac{e^{i l_{p}.\vec{p}}}{(\vec{p}^2+m_{1}^2)(\vec{q}^2+m_{1}^2)((\vec{p}+\vec{q})^2+m_{2}^2)}\nonumber\\
&+ \int \frac{d^d{p}}{(2\pi)^d}~\frac{d^{d}{q}}{(2\pi)^d}\sum_{l_{p}\neq 0}\frac{e^{i l_{p}.(\vec{p}+\vec{q})}}{(\vec{p}^2+m_{1}^2)(\vec{q}^2+m_{1}^2)((\vec{p}+\vec{q})^2+m_{2}^2)}\nonumber\\
& + \int \frac{d^d{p}}{(2\pi)^d}~\frac{d^{d}{q}}{(2\pi)^d}\sum_{\substack{\{{l_{p}},{l_{q}}\}\neq 0,\\ {l_{p}}\neq {l_{q}}}}\frac{e^{i l_{p}.\vec{p}+i l_{q}.\vec{q}}}{(\vec{p}^2+m_{1}^2)(\vec{q}^2+m_{1}^2)((\vec{p}+\vec{q})^2+m_{2}^2)}.
\end{align}
The first term appearing on the right hand side of Eq.~\eqref{sunin2} is given by the following relation \cite{ART000882346, Two-loop-two-point, Two-loop}:
\begin{align}\label{sunin4p}
 & \int \frac{d^d{p}}{(2\pi)^d}~\frac{d^{d}{q}}{(2\pi)^d}\frac{1}{(\vec{p}^2+m_{1}^2)(\vec{q}^2+m_{1}^2)((\vec{p}+\vec{q})^2+m_{2}^2)}\nonumber\\ 
 & = \frac{m_{2}^2}{2(4\pi)^4}\left(\frac{m_{2}^2}{4\pi}\right)^{-\epsilon}\frac{\Gamma^{2}\left(1+\frac{\epsilon}{2}\right)}{\left(1-\frac{\epsilon}{2}\right)\left(1-\epsilon\right)}\left[-\frac{2}{\epsilon^2}\left\{2\left(\frac{m^{2}_{1}}{m^{2}_{2}}\right)^{-\frac{\epsilon}{2}}+\left(2\frac{m^{2}_{1}}{m^{2}_{2}}-1\right)\left(\frac{m^{2}_{1}}{m^{2}_{2}}\right)^{-\epsilon}\right\}+\left\{\left(4\frac{m^{2}_{1}}{m^{2}_{2}}-1\right)^{1-\frac{\epsilon}{2}}\left(B[m^{2}_{2}, m^{2}_{1}]+O(\epsilon)\right)\right\}\right].
\end{align}
Using techniques from \cite{bijnens2014two, Sahu:2025bkp}, one can easily compute the second and the third term in Eq.~\eqref{sunin2}. They are given by:
\begin{align}
    &\int \frac{d^d{p}}{(2\pi)^d}~\frac{d^{d}{q}}{(2\pi)^d}\sum_{l_{p}\neq 0}\frac{e^{i l_{p}.\vec{p}}}{(\vec{p}^2+m_{1}^2)(\vec{q}^2+m_{1}^2)((\vec{p}+\vec{q})^2+m_{2}^2)}\nonumber\\
        & =\frac{1}{L^{2d-6}}\biggl\{ \frac{1}{16\pi^2}\left[\frac{2}{\epsilon}+\left(\log{4\pi}-\gamma_{E}-\log{m_{2}^2 L^2}\right)\right]\left(\frac{1}{4\pi}\theta^{(1)}\left(\frac{m_{1}^2 L^2}{4\pi}\right)+\frac{1}{m_{1}^2 L^2}\right)+ \frac{1}{(16\pi^2)^2}\int_{0}^{\infty}d\alpha\int_{0}^{\infty}d\beta \int_{0}^{\infty}d\gamma\frac{e^{-(\alpha+\beta)m_{1}^2L^2-\gamma m^{2}_{2}L^2}}{\mathcal{S}^{2}}\nonumber\\
   &\times \gamma\biggl(\left[(m^{2}_{2}L^2-m^{2}_{1}L^2)+2\frac{\beta-\gamma}{\mathcal{S}}\right] \left(\vartheta^{4}\left(\frac{(\beta+\gamma)}{4\pi\mathcal{S}}\right)-1\right)+(\beta-\gamma)\frac{(\beta+\gamma)}{4\pi\mathcal{S}^2}4~\vartheta^{3}\left(\frac{(\beta+\gamma)}{4\pi\mathcal{S}}\right)\vartheta'\left(\frac{(\beta+\gamma)}{4\pi\mathcal{S}}\right)\biggr)+O(\epsilon)\biggr\}. 
\end{align}
and\\
\begin{align}
    &\int \frac{d^d{p}}{(2\pi)^d}~\frac{d^{d}{q}}{(2\pi)^d}\sum_{l_{p}\neq 0}\frac{e^{i l_{p}.(\vec{p}+\vec{q})}}{(\vec{p}^2+m_{1}^2)(\vec{q}^2+m_{1}^2)((\vec{p}+\vec{q})^2+m_{2}^2)}\nonumber\\
    & = \int \frac{d^d{p}}{(2\pi)^d}~\frac{d^{d}{q}}{(2\pi)^d}\sum_{l_{p}\neq 0}\frac{e^{i l_{p}.\vec{p}}}{(\vec{p}^2+m_{2}^2)(\vec{q}^2+m_{1}^2)((\vec{p}+\vec{q})^2+m_{1}^2)}\nonumber\\
        & = \frac{1}{L^{2d-6}}\biggl\{\frac{1}{16\pi^2}\left[\frac{2}{\epsilon}+\left(\log{4\pi}-\gamma_{E}-\log{m_{1}^2 L^2}\right)\right]\left(\frac{1}{4\pi}\theta^{(1)}\left(\frac{m_{2}^2 L^2}{4\pi}\right)+\frac{1}{m_{2}^2 L^2}\right)+ \frac{1}{(16\pi^2)^2}\int_{0}^{\infty}d\alpha\int_{0}^{\infty}d\beta \int_{0}^{\infty}d\gamma\frac{e^{-\alpha m^{2}_{2}L^2-(\beta+\gamma)m^{2}_{1}L^2}}{\mathcal{S}^{3}}\gamma(\beta-\gamma)\nonumber\\
   & \times\left[2\left(\vartheta^{4}\left(\frac{(\beta+\gamma)}{4\pi\mathcal{S}}\right)-1\right)+\frac{(\beta+\gamma) }{4\pi\mathcal{S}}4~\vartheta^{3}\left(\frac{(\beta+\gamma)}{4\pi\mathcal{S}}\right)\vartheta'\left(\frac{(\beta+\gamma)}{4\pi\mathcal{S}}\right)\right]+O(\epsilon)\biggr\}.
   \end{align}
The last term in Eq.~\eqref{sunin2} is given by:
\begin{align}\label{sunxN3on}
    \int \frac{d^d{p}}{(2\pi)^d}~\frac{d^{d}{q}}{(2\pi)^d}\sum_{\substack{\{{l_{p}},{l_{q}}\}\neq 0,\\ {l_{p}}\neq {l_{q}}}}&\frac{e^{i l_{p}.\vec{p}+i l_{q}.\vec{q}}}{(\vec{p}^2+m_{1}^2)(\vec{q}^2+m_{1}^2)((\vec{p}+\vec{q})^2+m_{2}^2)}\nonumber\\
     = \frac{1}{L^{2d-6}}\biggl\{\frac{1}{(16\pi^2)^2}\int_{0}^{\infty}d\alpha\int_{0}^{\infty}d\beta\int_{0}^{\infty}d\gamma\frac{e^{-(\alpha+\beta)m^{2}_{1}L^2-\gamma m^{2}_{2}L^2}}{\mathcal{S}^{2}} &\biggl[\left(   \vartheta^{(2)}\left(0\, \biggr\rvert\,\frac{i }{4\pi \mathcal{S}}\begin{bmatrix}
   \beta+\gamma & -\gamma\\
   -\gamma & \alpha+\gamma
   \end{bmatrix}\right)\right)^{4}-\vartheta^{4}\left(\frac{(\alpha+\beta)}{4\pi\mathcal{S}}\right)-\vartheta^{4}\left(\frac{(\beta+\gamma)}{4\pi\mathcal{S}}\right)-\vartheta^{4}\left(\frac{(\gamma+\alpha)}{4\pi\mathcal{S}}\right)+2\biggr]\nonumber\\
   & +O(\epsilon)\biggr\},\nonumber\\
\end{align}\\

where $\mathcal{S}=\alpha\beta+\beta\gamma+\gamma\alpha$ in the above relations and $\vartheta^{(g)}(z|\tau)= \sum_{n\in Z^{g}}e^{2\pi i (\frac{1}{2} n^{T}.\tau.n+n^{T}.z)}$ is the Riemann-theta function.\\

Now we show how to compute the bubble diagrams emerging at two loops (the diagrams appearing in $\mathcal{I}_{3}$, $\mathcal{I}_{4}$, $\mathcal{I}_{5}$) . Here we only give the calculation for the inhomogenous bubble diagram (the diagram appearing in $\mathcal{I}_{5}$), from which one can easily compute the homogenous ones, putting $m_{2}=m_{1}$ or following \cite{Sahu:2025bkp}. Thus using Eq.~\eqref{discrete}, in $d=4-\epsilon$, the inhomogeneous bubble term is given by:
\begin{align}
    &\frac{1}{L^{2d}} \left(\sum_{p\neq 0}\frac{1}{(\vec{p}^2+m^{2}_{1})}\right) \left(\sum_{q\neq 0}\frac{1}{(\vec{q}^2+m^{2}_{2})}\right)\nonumber\\
    & = \frac{1}{L^{2d-4}}\biggl[\frac{1}{16\pi^2}\theta^{(1)}\left(\frac{m_{1}^2 L^2}{4\pi}\right)\theta^{(1)}\left(\frac{m_{2}^2 L^2}{4\pi}\right)+\frac{1}{4\pi}\theta^{(1)}\left(\frac{m_{2}^2 L^2}{4\pi}\right)\frac{m_{1}^2 L^2}{(4\pi)^2}\left(\frac{-2}{\epsilon}+\gamma_{E} -1+\log{\left(\frac{m_{1}^2 L^2}{4\pi}\right)}\right)\nonumber\\
    & +\frac{1}{4\pi}\theta^{(1)}\left(\frac{m_{1}^2 L^2}{4\pi}\right)\frac{(m_{2}^2 L^2)}{(4\pi)^2}\left(\frac{-2}{\epsilon}+\gamma_{E} -1+\log{\left(\frac{(m_{2}^2 L^2)}{4\pi}\right)}\right)+\frac{(m^{2}_{1}L^2)(m^{2}_{2}L^2)}{(16\pi^2)^2}\biggl(\frac{-2}{\epsilon}+\gamma_{E} -1+\log{\left(\frac{m_{2}^2 L^2}{4\pi}\right)}+\frac{\epsilon}{24}\biggl(6\gamma_{E}^2+\pi^2+12\nonumber\\
    &-12\gamma_{E}+6\log^{2}{\left(\frac{m^{2}_{2}L^2}{4\pi}\right)}+12(\gamma_{E}-1)\log{\left(\frac{m^{2}_{2}L^2}{4\pi}\right)}\biggr)+O(\epsilon^2)\biggr)\biggl(\frac{-2}{\epsilon}+\gamma_{E} -1+\log{\left(\frac{m_{1}^2 L^2}{4\pi}\right)}+\frac{\epsilon}{24}\biggl(6\gamma_{E}^2+\pi^2+12\nonumber\\
    &-12\gamma_{E}+6\log^{2}{\left(\frac{m^{2}_{1}L^2}{4\pi}\right)}+12(\gamma_{E}-1)\log{\left(\frac{m^{2}_{1}L^2}{4\pi}\right)}\biggr)+O(\epsilon^2)\biggr)+O(\epsilon)\biggr]
\end{align}
\section{List of functions}\label{appenB}
We give here the list of functions used in Eqs.~\eqref{ratef v} and \eqref{ratef iv}:
    \begin{align}
        I_{1}[\tilde{m}^{2}] &=\frac{3}{(16\pi^2)^2}\int_{0}^{\infty}d\alpha\int_{0}^{\infty}d\beta \int_{0}^{\infty}d\gamma\frac{e^{-(\alpha+\beta+\gamma)\tilde{m}^2 }}{\mathcal{S}^{3}}\gamma(\beta-\gamma)\left[2\left(\vartheta^{4}\left(\frac{(\beta+\gamma)}{4\pi\mathcal{S}}\right)-1\right)+4\frac{(\beta+\gamma) }{4\pi\mathcal{S}}~\vartheta^{3}\left(\frac{(\beta+\gamma)}{4\pi\mathcal{S}}\right)\vartheta'\left(\frac{(\beta+\gamma)}{4\pi\mathcal{S}}\right)\right],\\
        I_{2}[\tilde{m}^{2}] &=\frac{1}{(16\pi^2)^2}\int_{0}^{\infty}d\alpha\int_{0}^{\infty}d\beta\int_{0}^{\infty}d\gamma\frac{e^{-(\alpha+\beta+\gamma)\tilde{m}^2 }}{\mathcal{S}^{2}} \biggl[ \left(\vartheta^{(2)}\left(0\, \biggr\rvert\,\frac{i}{4\pi \mathcal{S}}\begin{bmatrix}
   \beta+\gamma & -\gamma\\
   -\gamma & \alpha+\gamma
   \end{bmatrix}\right)\right)^{4}-\vartheta^{4}\left(\frac{(\alpha+\beta)}{4\pi\mathcal{S}}\right)\nonumber\\
    &-\vartheta^{4}\left(\frac{(\beta+\gamma)}{4\pi\mathcal{S}}\right)-\vartheta^{4}\left(\frac{(\gamma+\alpha)}{4\pi\mathcal{S}}\right)+2\biggr],\\
    K_{1}[\tilde{m}_{2}^{2}, \tilde{m}^{2}_{1}] & =  \frac{1}{(16\pi^2)^2}\int_{0}^{\infty}d\alpha\int_{0}^{\infty}d\beta \int_{0}^{\infty}d\gamma\frac{e^{-(\alpha+\beta)\tilde{m}_{1}^2-\gamma \tilde{m}^{2}_{2}}}{\mathcal{S}^{2}}\gamma\biggl(\left[(\tilde{m}^{2}_{2}-\tilde{m}^{2}_{1})+2\frac{\beta-\gamma}{\mathcal{S}}\right]\left(\vartheta^{4}\left(\frac{(\beta+\gamma)}{4\pi\mathcal{S}}\right)-1\right)\nonumber\\
    & +(\beta-\gamma)\frac{(\beta+\gamma)}{4\pi\mathcal{S}^2}4~\vartheta^{3}\left(\frac{(\beta+\gamma)}{4\pi\mathcal{S}}\right)\vartheta'\left(\frac{(\beta+\gamma)}{4\pi\mathcal{S}}\right)\biggr),\\
     K_{2}[\tilde{m}_{2}^{2}, \tilde{m}^{2}_{1}] & = \frac{1}{(16\pi^2)^2}\int_{0}^{\infty}d\alpha\int_{0}^{\infty}d\beta \int_{0}^{\infty}d\gamma\frac{e^{-\alpha \tilde{m}^{2}_{2}-(\beta+\gamma)\tilde{m}^{2}_{1}}}{\mathcal{S}^{3}}\gamma(\beta-\gamma)\left[2\left(\vartheta^{4}\left(\frac{(\beta+\gamma)}{4\pi\mathcal{S}}\right)-1\right)+\frac{(\beta+\gamma) }{4\pi\mathcal{S}}4~\vartheta^{3}\left(\frac{(\beta+\gamma)}{4\pi\mathcal{S}}\right)\vartheta'\left(\frac{(\beta+\gamma)}{4\pi\mathcal{S}}\right)\right],\\ 
     K_{3}[\tilde{m}_{2}^{2}, \tilde{m}^{2}_{1}] & = \frac{1}{(16\pi^2)^2}\int_{0}^{\infty}d\alpha\int_{0}^{\infty}d\beta\int_{0}^{\infty}d\gamma\frac{e^{-(\alpha+\beta)\tilde{m}^{2}_{1}-\gamma \tilde{m}^{2}_{2}}}{\mathcal{S}^{2}} \biggl[   \left(\vartheta^{(2)}\left(0\, \biggr\rvert\,\frac{i}{4\pi \mathcal{S}}\begin{bmatrix}
   \beta+\gamma & -\gamma\\
   -\gamma & \alpha+\gamma
   \end{bmatrix}\right)\right)^{4}-\vartheta^{4}\left(\frac{(\alpha+\beta)}{4\pi\mathcal{S}}\right)\nonumber\\
   &-\vartheta^{4}\left(\frac{(\beta+\gamma)}{4\pi\mathcal{S}}\right)-\vartheta^{4}\left(\frac{(\gamma+\alpha)}{4\pi\mathcal{S}}\right)+2\biggr],\\
    \Delta(\tilde{m}^{2})&=\theta(\tilde{m}^{2})-\theta(0)=\int_0^\infty d\sigma \frac{1-e^{-\sigma \tilde{m}^2}}\sigma\left(\vartheta^d(\sigma)-1-\sigma^{-d/2}\right),\\
    A&=-\frac{2}{\sqrt{3}}\int_{0}^{\frac{\pi}{3}}dy~\log{\left(2 \sin{\frac{y}{2}}\right)},\nonumber\\
    B[\tilde{m}^{2}_{2}, \tilde{m}^{2}_{1}]&=-2\int_{0}^{r}dy~\log{\left(2 \sin{\frac{y}{2}}\right)}\\
    \Delta^{(\epsilon)}(\tilde{m}^{2})&=\int_{0}^{\infty}d\sigma \frac{1-e^{-\tilde{m}^{2}\sigma}}{\sigma}\left(\vartheta^{4}(\sigma)\log{\vartheta(\sigma)}+\frac{\log{\sigma}}{2\sigma^{2}}\right).\\
    \end{align}\\
    With $\mathcal{S}=\alpha\beta+\beta\gamma+\gamma\alpha$ and $r=\cos^{-1}\left(1-\frac{\tilde{m}^2_{2}}{2\tilde{m}^{2}_{1}}\right)$.\\
    
      Here the function $\Delta^{(\epsilon)}$ is the term that appears at the first order of the $\epsilon$-expansion of the function $\Delta$ (see I). The functions $\theta(\tilde{m}^{2})$, $\theta^{(1)}(\tilde{m}^{2})$, $\theta^{(2)}(\tilde{m}^{2})$ are the same as defined in appendix \ref{appenA}. The integrals $I_{1}[\tilde{m}^{2}]$ $I_{1}[\tilde{m}^{2}]$, $ K_{1}[\tilde{m}_{2}^{2}, \tilde{m}^{2}_{1}]$, $ K_{2}[\tilde{m}_{2}^{2}, \tilde{m}^{2}_{1}]$ and $ K_{3}[\tilde{m}_{2}^{2}, \tilde{m}^{2}_{1}]$  are rather demanding to compute. Numerical integration of these integrals are highly non-trivial. Numerical implementation of them follows from \cite{bijnens2014two}.
      
\twocolumngrid
\bibliography{main}
\bibliographystyle{apsrev4-1} 
    
\end{document}